\documentclass[]{interact}

\usepackage{epstopdf}
\usepackage[caption=false]{subfig}

\usepackage[numbers,sort&compress]{natbib}
\bibpunct[, ]{[}{]}{,}{n}{,}{,}
\renewcommand\bibfont{\fontsize{10}{12}\selectfont}

\usepackage[french,english]{babel}
\usepackage{float}
\usepackage{url}
\usepackage{color}
\usepackage[normalem]{ulem}

\theoremstyle{plain}

\theoremstyle{definition}

\theoremstyle{remark}

\newcommand{\ActivationEnergy}{\mathfrak{E}}
\newcommand{\Avogadro}{\mathcal{N}_{\textsc{a}}}
\newcommand{\Boltzmann}{k_{\textsc{b}}}
\newcommand{\ChemK}{\mathcal{K}}
\newcommand{\Mass}{\mathfrak{m}}
\newcommand{\MoleMass}{W}
\newcommand{\n}{n}
\newcommand{\p}{p}
\newcommand{\RU}{R}
\newcommand{\T}{T}
\newcommand{\Vel}{\mathcal{V}}

\begin{document}

\articletype{ORIGINAL ARTICLE}

\title{Importance of mass and enthalpy conservation in the modeling of titania nanoparticles flame synthesis}

\author{
\name{Jean-Maxime Orlac'h\textsuperscript{a}\thanks{CONTACT J.-M. Orlac'h. Now at ONERA, 29 avenue de la Division Leclerc, 92320 Châtillon, France. Email: jean-maxime.orlach@onera.fr}, Nasser Darabiha\textsuperscript{a}, Vincent Giovangigli\textsuperscript{b} and Benedetta~Franzelli\textsuperscript{a}}
\affil{\textsuperscript{a}Laboratoire EM2C, CNRS, Ecole CentraleSup\'{e}lec, 8-10 rue Joliot Curie, 91190 Gif-sur-Yvette, France \\
\textsuperscript{b}Centre de Math\'{e}matiques Appliqu\'{e}es, CNRS, Ecole Polytechnique, 91128 Palaiseau, France}
}

\maketitle

\begin{abstract}
In most simulations of fine particles in reacting flows, including sooting flames, enthalpy exchanges between gas and particle phases and differential diffusion between the two phases are most often neglected, since the particle mass fraction is generally very small. However, when the nanoparticles mass fraction is very large representing up to 50 \% of the mixture mass, the conservation of the total enthalpy and/or the total mass becomes critical. In the present paper, we investigate the impact of mass and enthalpy conservation in the modeling of titania nanoparticles synthesis in flames, classically characterized by a high conversation rate and consequently a high nanoparticles concentration. It is shown that when the nanoparticles concentration is high, neglecting the enthalpy of the particle phase may lead to almost 70\,\% relative error on the temperature profile and to relative errors on the main titania species mass fractions and combustion products ranging from 20\,\% to 100\,\%. It is also established that neglecting the differential diffusion of the gas phase with respect to the particle phase is also significant, with almost 15\,\% relative error on the TiO$_{2}$ mole fraction, although the effect on combustion products is minor.

\end{abstract}


\section{Introduction}

Flame processes are widely used for the manufacture of several nano-structured commodities, such as titanium dioxide -- titania --, carbon black and fumed silica \cite{pratsinis_flame_1998,xu_simultaneous_2015,kelesidis_flame_2017,schulz_gas-phase_2018}, representing a billion dollar industry. Nanoparticle synthesis requires a fine control of the particle size and shape distribution, and of the nanoparticle crystal phase with desired properties depending on the targeted application. Therefore, detailed and accurate modeling is of critical importance for the optimization of nanoparticle production in flame reactors \cite{weise_buoyancy_2013,raman_modeling_2016,sellmann_detailed_2018}.

Titanium dioxide is used as a white pigment -- e.g. in paintings, solar creams, cosmetics -- as a catalyst support, and as a photocatalyst. Even if in most laboratory-scale experimental studies of titania nanoparticles flame synthesis the precursor concentration generally represents a few percent of the oxidizer flow rate \cite{george_formation_1973,pratsinis_role_1996,chow_flame_1996}, industrial aerosol reactors are usually operated at high precursor mass fraction, possibly more than 50\% of the oxidizer flow rate \cite{pratsinis_kinetics_1990,heine_agglomerate_2007}. Since the reaction yield is rather high, as up to 50\% of the injected precursor can be converted into TiO$_{2}$ particles \cite{chow_flame_1996}, the mass fraction of the particle phase can be non-negligible in comparison to that of the gas phase. This yields a strong coupling between both phases, such that exchanges of mass and energy between the two phases will no longer be negligible. Other types of nanoparticles are also concerned with high conversion yield and high concentration, such as SiO$_{2}$ nanoparticles produced from SiCl$_{4}$ \cite{hwang_measurements_2001,kim_modeling_2003, lee_simulation_2001,morgan_modelling_2005} or HMDSO \cite{grohn_design_2011,chrystie_comparative_2018}. In the present work, we focus on titania nanoparticles as a representative test case.

The modeling of metal oxide nanoparticle synthesis in flames is the object of a continued interest \cite{raman_modeling_2016}. In particular, a strong effort has been dedicated over the past decades to the development of accurate and efficient numerical models for titania nanoparticles synthesis in flames and reacting flows. As a first approach, only the particle phase can be accounted for while prescribing the experimental temperature profiles. In that case, no equations are solved for enthalpy, fluid flow velocity, or gaseous species mass fractions, and the state of the gas corresponding to experimental measurements is taken as an input to the nanoparticle model. Therefore, it is not necessary to model self-consistently the gas phase to ensure conservation of the mixture mass and enthalpy. For this, 0D reactors \cite{pratsinis_competition_1998,spicer_titania_2002,tsantilis_narrowing_2004,heine_agglomerate_2007,morgan_new_2006} or arrays of 0D reactor models are considered \cite{boje_detailed_2017,west_detailed_2009,mehta_multiscale_2010, mehta_role_2013}. However, such approaches require experimental data on the temperature, which are not always available.

Some studies have accounted for a two-way coupling between gas and particle phases by incorporating in the gas phase a pseudo gas component representing the overall nanoparticle mass fraction \cite{johannessen_computational_2001}, thereby ensuring the global conservation of enthalpy and mass. Other authors have employed fully two-way coupled models, without any pseudo gas component \cite{wang_modeling_2005,garrick_modeling_2011,akroyd_coupled_2011,xu_cfd-population_2017}. Among them, Wang and Garrick neglected enthalpy and mass exchanges between the gas and particle phases \cite{wang_modeling_2005,garrick_modeling_2011}, while other authors \cite{akroyd_coupled_2011,xu_cfd-population_2017} solved the balance equation for the enthalpy of the gas phase accounting for an energy exchange term between the gas and particle phases to ensure global energy conservation. Yet, to our knowledge no one has ever quantified the impact of mass and enthalpy conservation for such applications.

As such, two main formulations can be found in the literature for the diffusion velocities and enthalpy conservation equation: a conservative one and a non-conservative one. The conservative formulation is such that the total mass and enthalpy of the mixture are both conserved, while the non-conservative formulation neglects the enthalpy contribution and the differential diffusion of the nanoparticles. For low aerosol mass fractions, e.g. soot particles, both models are equivalent in practice \cite{zimmer_investigation_2017} since the nanoparticles mass fraction generally remains low \cite{smooke_investigation_2004,wang_formation_2011,franzelli_time-resolved_2015}, and the conservation of mass and enthalpy is not an issue. Therefore, both kinds of models can be found in that case, namely non-conservative \cite{wang_modeling_2005,garrick_modeling_2011} and conservative ones \cite{akroyd_coupled_2011,xu_cfd-population_2017}, whereas some authors do not detail the approach they actually follow in their codes \cite{eaves_coflame:_2016,mueller_joint_2009,bisetti_formation_2012}.

In the present paper, we demonstrate the importance of conservation of mass and enthalpy for highly concentrated aerosol reactors. Also, we analyze the effect of neglecting the contribution of the particle phase to the mixture enthalpy or to the mass diffusion on numerical simulations of nanoparticle synthesis in flames. In addition, we show that the use of non-conservative models can yield numerical instabilities. 

For this, we consider one-dimensional laminar premixed and counterflow methane-oxygen-TiCl$_{4}$ flames. Such idealized flame configurations are the simplest possible, so that the conservation of enthalpy or mass can be easily assessed. Besides, as many turbulent flame models rely on preliminary one-dimensional calculations on idealized cases, accurately modeling one-dimensional (1D) premixed and counterflow flames is a necessary step before adressing the modeling of turbulent flames.

A classical one-step nucleation kinetics is used to describe titania formation \cite{pratsinis_kinetics_1990}. Although very simple in essence, this scheme has been validated against experimental data previously in several publications \cite{xiong_formation_1993,pratsinis_competition_1998,spicer_titania_2002,tsantilis_narrowing_2004,morgan_new_2006,heine_agglomerate_2007,xu_cfd-population_2017}. This simple scheme reproduces well the precursor consumption rate and the global titania production rate, so it is well suited for the present study.

The paper is organized as follows. In section \ref{sec:dispphaseeq}, the conservative and non-conservative formulations are detailed. In section \ref{sec:applications}, we present the test cases adopted here, namely titania nanoparticle production from TiCl$_{4}$ in 1D premixed and non-premixed laminar flames. In section \ref{sec:1dp} the conservation of enthalpy is studied in premixed flames. Finally, in section \ref{sec:1dcf}, the effects of enthalpy conservation and differential diffusion on 1D non-premixed flames are considered.

\section{A detailed conservative model for flame synthesis of nanoparticles}
\label{sec:dispphaseeq}

The conservation equations are detailed in subsection \ref{subsec:conseq}. Then, we state the expressions for the transport fluxes and mixture enthalpy: we will present first in subsection \ref{subsec:diffvel} the two formulations -- conservative and non-conservative -- for the gaseous species and particles diffusion velocities. Next, in subsection \ref{subsec:enthalpy} the two formulations for the enthalpy conservation equation are stated.

It is worth mentioning that when high particle volume fractions are encountered, the dynamics of nanoparticles can be different compared to dilute aerosols \cite{heine_agglomerate_2007, buesser_coagulation_2009}. Even at high aerosol mass fractions, the nanoparticles volume fraction remains low in general because the density of each individual nanoparticle is much larger than the average gas density. However, high level of fractality can have a strong effect on the nanoparticles dynamics as it increases the effective volume fraction occupied by the aerosol. It can in particular significantly affect the collision frequencies, or even lead to gelation of the aerosol \cite{heine_agglomerate_2007}. Here, for the sake of simplicity, we neglect the nanoparticles fractality, so that the effective particle volume fraction remains low, and the aerosol remains in the dilute regime. In that case, the aerosol General Dynamic Equation (GDE) remains valid. However, the study conducted here could be easily extended to any kind of aerosol kinetic equation, provided that such an equation is known.

\subsection{Conservation equations}
\label{subsec:conseq}
In this subsection we first write the general continuous formulation, and then the discrete formulation of the sectional method.

\subsubsection{Continuous formulation}

We consider here that the two phases are in thermal and mechanical equilibrium, so that the velocities of the gas -- subscript g -- and solid particle -- subscript p -- phases are equal: $\boldsymbol{u}_{\text{g}} = \boldsymbol{u}_{\text{p}} = \boldsymbol{u}$,  where $\boldsymbol{u}$ is the mixture-averaged velocity, and their temperatures are identical: $\T_{\text{g}} = \T_{\text{p}} = \T$, where $\T$ is the mixture temperature. These assumptions are generally made in the modeling of fine particle transport.

This allows to consider the mixture as a unique dispersed phase, whose components can be either gas-phase molecular species or solid-state nanoparticles. In this so-called 'one-mixture' model, the classical equations for conservation of mass and enthalpy of a multicomponent gas mixture are retained, but the mixture density and enthalpy must now include both the gas and particle phases contributions. The mixture density reads $\rho = \rho_{\text{g}} + \rho_{\text{p}}$, where $\rho_{\text{g}}$ and $\rho_{\text{p}}$ are the mass densities of the gas and solid particle mixtures, respectively:
\begin{gather}
 \rho_{\text{g}} = \sum_{k=1}^{N_{\text{g}}} \rho_{k}, \label{eq:Rhog} \\
 \rho_{\text{p}} = \int \rho_{\text{s}} \, q(v) \, \mathrm{d}v. \label{eq:RhopCont}
\end{gather}
where $\rho_{k}$ is the k$^{\text{th}}$ species density, $N_{\text{g}}$ denotes the number of gas-phase species, and $\rho_{\text{s}}$ is the density of a solid particle (supposed equal for each particle). The internal variable $v$ represents the nanoparticle volume, $q(v)=v n(v)$ (in cm$^{-3}$) is the volume density of the aerosol where $n(v)$ is the number density distribution.

The $k^{\text{th}}$ species mass fraction $Y_{k}$ and the mass fraction $Y(v) \, \mathrm{d}v$ of nanoparticles whose volume lies in the range $[v,v+\mathrm{d}v]$ are respectively defined as:
\begin{gather}
 Y_{k} = \frac{\rho_{k}}{\rho}, \quad k=1, \ldots, N_{\text{g}}, \label{eq:OneMixtYk} \\
 Y(v) = \frac{\rho_{\text{s}} \, q(v) }{\rho}. \label{eq:OneMixtYv}
\end{gather}
The global mass conservation in the mixture implies that the mass fractions sum to unity:
\begin{equation}
Y_{\text{g}} + Y_{\text{p}} = 1,
\label{eq:OneMixtSumYUnity}
\end{equation}
where $Y_{\text{g}}$ and $Y_{\text{p}}$ are the total gas and particle mass fractions, respectively:
\begin{gather}
Y_{\text{g}}=\sum_{k=1}^{N_{\text{g}}} Y_{k}, \\
Y_{\text{p}}=\int Y(v) \, \mathrm{d}v. \label{eq:YpCont}
\end{gather}

The particles volume fraction is then given by:
\begin{equation}
f_{\textsc{v}} = \int q(v) \, \mathrm{d} v = \frac{\rho Y_{\text{p}}}{\rho_{\text{s}}}.
\end{equation}

The conservation equations of mass, species and particle mass fractions then read:
\begin{gather}
\frac{\partial \rho}{\partial t} + \boldsymbol{\nabla} \cdot (\rho \boldsymbol{u}) = 0, \label{eq:OneMixtMassCons} \\
\frac{\partial (\rho Y_{k})}{\partial t} + \boldsymbol{\nabla} \cdot (\rho Y_{k} \boldsymbol{u} + \rho Y_{k} \boldsymbol{\Vel}_{k}) = \MoleMass_{k} \dot{\omega}_{k}, \quad k=1, \ldots, N_{\text{g}}, \label{eq:OneMixtSpeciesCons} \\
\frac{\partial (\rho Y(v))}{\partial t} + \boldsymbol{\nabla} \cdot (\rho Y(v) \boldsymbol{u} + \rho Y(v) \boldsymbol{\Vel}(v)) = \rho_{\text{s}} \, \dot{q}(v), v \in ]0, + \infty[, \label{eq:OneMixtContParticleCons}
\end{gather}
where $\MoleMass_{k}$ is the $k^{\text{th}}$ species molar mass, $\dot{\omega}_{k}$ is the $k^{\text{th}}$ species molar production rate, and $\boldsymbol{\Vel}_{k}$ is the $k^{\text{th}}$ species diffusion velocity. $\boldsymbol{\Vel}(v)$ is the diffusion velocity of particles whose volume lies in the range $[v,v+\mathrm{d}v]$, and $\dot{q}(v)$ is the volumetric particle source term. Eq. \eqref{eq:OneMixtContParticleCons} is merely the General Dynamic Equation (GDE) for aerosols \cite{friedlander_smoke_2000}. The mass exchange between the phases imposes:
\begin{equation}
\sum_{k=1}^{N_{\text{g}}} \MoleMass_{k} \dot{\omega}_{k} + \int \rho_{\text{s}} \, \dot{q}(v) \, \mathrm{d}v = 0. \label{eq:SourceConsCont}
\end{equation}
Therefore, to ensure mass conservation, the diffusive fluxes must sum to zero:
\begin{equation}
\sum_{k=1}^{N_{\text{g}}} \rho Y_{k} \boldsymbol{\Vel}_{k}+ \int \rho Y(v) \boldsymbol{\Vel}(v) \, \mathrm{d}v = 0,
\label{eq:OneMixtContDiffFluxCons}
\end{equation}
so that the constraint Eq. \eqref{eq:OneMixtSumYUnity} is satisfied \cite{giovangigli_mass_1990}. The system is closed by the perfect gas law $\rho = \p \overline{\MoleMass}/(R \T)$, where $\p$ is the pressure, $R$ is the universal constant, and $\overline{\MoleMass}$ is the mean molar mass of the mixture, given by
\begin{equation}
\frac{1}{\overline{\MoleMass}} = \sum_{k=1}^{N_{\text{g}}} \frac{Y_{k}}{\MoleMass_{k}} + \int \frac{Y(v)}{\MoleMass(v)} \, \mathrm{d}v, \label{eq:MoleMassCont}
\end{equation}
where $\MoleMass(v) \, \mathrm{d}v = \rho_{\text{s}} \, v \, \Avogadro \, \mathrm{d}v$ is the molar mass of the particles whose volume lies in the range $[v,v+\mathrm{d}v]$, with $\Avogadro$ the Avogadro number. Note that if one neglects the nanoparticles contribution to the mixture density, the above perfect gas law can be replaced with a non-conservative perfect gas law $\rho = \p \overline{\MoleMass_{\text{g}}}/(R \T)$, where $\overline{\MoleMass}_{\text{g}}$ is the average gas molar mass, given by
\begin{equation}
\frac{Y_{\text{g}}}{\overline{\MoleMass}_{\text{g}}} = \sum_{k=1}^{N_{\text{g}}} \frac{Y_{k}}{\MoleMass_{\text{k}}}.
\end{equation}

The momentum equation reads
\begin{equation}
\frac{\partial (\rho \boldsymbol{u})}{\partial t} + \boldsymbol{\nabla} \cdot ( \rho \boldsymbol{u} \boldsymbol{u} ) + \boldsymbol{\nabla} p + \boldsymbol{\nabla} \cdot \boldsymbol{\Pi} = 0, \label{eq:OneMixtMomentumCons}
\end{equation}
where $p$ is the pressure and $\boldsymbol{\Pi}$ is the viscous tensor of the -- gas and particles -- mixture.

Finally, the -- isobaric -- conservative balance equation for the mixture enthalpy can be written as:
\begin{equation}
\frac{\partial (\rho h)}{\partial t} + \boldsymbol{\nabla} \cdot (\rho h \boldsymbol{u}) + \boldsymbol{\nabla} \cdot \boldsymbol{Q} = 0, \label{eq:OneMixtContEnthalpyCons}
\end{equation}
where $h$ is the mixture specific enthalpy, and $\boldsymbol{Q}$ is the heat flux.

\subsubsection{Discrete formulation}

When a sectional model is used, the nanoparticle volume space is no longer continuous but is discretized into a finite number of sections: $N_{\text{s}}$. The conservative model described in the previous subsection is easily adapted to a discrete formulation. The corresponding equations are detailed below.

\par

In the discrete formulation, the mass density of the solid particle mixture (Eq. \eqref{eq:RhopCont}) reads:
\begin{equation}
\rho_{\text{p}} = \sum_{i=1}^{N_{\text{s}}} \rho_{i},
\label{eq:RhopDisc}
\end{equation}
where $\rho_{i}$ is the $i^{\text{th}}$ section density given by:
\begin{equation}
\rho_{i} = \rho \int_{i} Y(v) \, \mathrm{d}v, \quad i=1, \ldots, N_{\text{s}}.
\label{eq:RhoiDisc}
\end{equation}
The integration in Eq. \eqref{eq:RhoiDisc} is over the $i^{\text{th}}$ section, i.e. $\int_{i} \ldots = \int_{v_{i}^{\text{min}}}^{v_{i}^{\text{max}}} \ldots$ where $v_{i}^{\text{min}}$ and $v_{i}^{\text{max}}$ are the minimum and maximum volumes of the $i^{\text{th}}$ section. The $i^{\text{th}}$ section nanoparticle mass fraction is then defined as:
\begin{equation}
 Y_{i} = \int_{i} Y(v) \, \mathrm{d} v = \frac{\rho_{i}}{\rho}, \quad i=1, \ldots, N_{\text{s}}.
\label{eq:OneMixtYi}
\end{equation}
The total particle mass fraction (Eq. \eqref{eq:YpCont}) reads:
\begin{equation}
Y_{\text{p}} = \sum_{i=1}^{N_{\text{s}}} Y_{i} = \frac{\rho_{\text{p}}}{\rho} = \frac{\rho_{\text{s}} f_{\textsc{v}}}{\rho}.
\end{equation}
The $i^{\text{th}}$ section volume fraction $Q_{i}$ is expressed as:
\begin{equation}
Q_{i} = \frac{\rho}{\rho_{\text{s}}} Y_{i}.
\end{equation}

The conservation equation for the $i^{\text{th}}$ section nanoparticle mass fraction (Eq. \eqref{eq:OneMixtContParticleCons}) then reads:
\begin{equation}
\frac{\partial (\rho Y_{i})}{\partial t} + \boldsymbol{\nabla} \cdot (\rho Y_{i} \boldsymbol{u} + \rho Y_{i} \boldsymbol{\Vel}_{i}) = \rho_{\text{s}} \dot{Q}_{i}, \quad i=1, \ldots, N_{\text{s}}, \label{eq:OneMixtDiscParticleCons}
\end{equation}
where $Y_{i} \boldsymbol{\Vel}_{i} = \int_{i} Y(v) \boldsymbol{\Vel}(v) \, \mathrm{d} v$ is the diffusion velocity of particles of the $i^{\text{th}}$ section, and $\dot{Q}_{i} = \int_{i} \dot{q}(v) \, \mathrm{d} v$ is the particle mass source term for the $i^{\text{th}}$ section.

The mass conservation (Eq. \eqref{eq:SourceConsCont}) now reads:
\begin{equation}
\sum_{k=1}^{N_{\text{g}}} \MoleMass_{k} \dot{\omega}_{k} + \sum_{i=1}^{N_{\text{s}}} \rho_{\text{s}} \dot{Q}_{i} = 0,
\end{equation}
As in the continuous case, the diffusive fluxes must sum to zero:
\begin{equation}
\sum_{k=1}^{N_{\text{g}}} \rho Y_{k} \boldsymbol{\Vel} _{k}+ \sum_{i=1}^{N_{\text{s}}} \rho Y_{i} \boldsymbol{\Vel} _{i}= 0,
\label{eq:OneMixtDiscDiffFluxCons}
\end{equation}
so that the constraint Eq. \eqref{eq:OneMixtSumYUnity} is satisfied \cite{giovangigli_mass_1990}. As well, the mean molar mass of the mixture (Eq. \eqref{eq:MoleMassCont}) is now given by
\begin{equation}
\frac{1}{\overline{\MoleMass}} = \sum_{k=1}^{N_{\text{g}}} \frac{Y_{k}}{\MoleMass_{k}} + \sum_{i=1}^{N_{\text{s}}} \frac{Y_{i}}{\MoleMass_{i}},
\end{equation}
where $\MoleMass_{i}$ is the mean molar mass of particles in the $i^{\text{th}}$ section.

Finally, the momentum equation (Eq. \eqref{eq:OneMixtMomentumCons}) and the enthalpy conservation equation (Eq. \eqref{eq:OneMixtContEnthalpyCons}) remain unchanged in the discrete case.

In practice, we assume classically \cite{gelbard_sectional_1980} that inside each section $i$, whose volume lies in the range $[v_{i}^{\text{min}},v_{i}^{\text{max}}]$, the particle volume fraction density $q(v)=v \, n(v)$ is constant and equal to $q_{i}$. The particle size distribution discretization follows a geometric progression.	The last section can be considered as a "trash" section which contains very big unexpected particles from $v^{\text{MAX}}$ to $v^{\text{BIG}}$ and guarantees particles mass conservation. The value of $v^{\text{BIG}}$ is chosen as an unattainable particle volume. The value of $v^{\text{MAX}}$ corresponds to a characteristic volume of the expected biggest particles and is chosen as the maximum particle volume resolved accurately. Supposing that the particles are spherical, the surface of a particle of volume $v$ reads $s(v) = \pi^{1/3} (6 v)^{2/3}$, and the diameter reads $d_{\text{p}}(v) = (6 v /\pi)^{1/3}$.

\subsection{Diffusion velocities}
\label{subsec:diffvel}
We detail here the two formulations used in the literature for the diffusion velocities, a conservative and a non-conservative one.

In the conservative formulation, the diffusion velocities of the gas-phase species and nanoparticles are given by \cite{eaves_coflame:_2016}:
\begin{gather}
\boldsymbol{\Vel} _{k} = - D_{k} \, \boldsymbol{\nabla} \ln{X_{k}} + \boldsymbol{u}_{\text{cor}}, \label{eq:OneMixtDiscDiffFluxGasCons} \\
\boldsymbol{\Vel} _{i}= \boldsymbol{v}_\text{th} - D_{i} \boldsymbol{\nabla} \ln{Y_{i}} + \boldsymbol{u}_{\text{cor}}. \label{eq:OneMixtDiscDiffFluxPartCons}
\end{gather}
where $D_{k}$ is the $k^{\text{th}}$ species mixture-averaged diffusion coefficient, $X_{k}$ is the $k^{\text{th}}$ species mole fraction, $\boldsymbol{v}_\text{th}$ is the thermophoretic velocity \cite{derjaguin_experimental_1966}, $D_{i}$ is the diffusion coefficient of particles of the $i^{\text{th}}$ section, and $\boldsymbol{u}_{\text{cor}}$ is a correction velocity, which is taken such that the constraint Eq. \eqref{eq:OneMixtDiscDiffFluxCons} is satisfied.

The particles diffusion coefficients are classically expressed as in the \textit{free molecular} regime \cite{epstein_resistance_1924,friedlander_smoke_2000}:
\begin{equation}
D_{i} = \frac{\Boltzmann T}{\left(1+\frac{\alpha_{\textsc{t}}\pi}{8}\right) \frac{\pi}{3}n \Mass \bar{c}d_{i}^{2}},
\end{equation}
where $\Mass$ is the average mass of a gas particle, $n$ is the gas number density, $\bar{c} = \sqrt{\frac{8 \Boltzmann \T}{\pi \Mass}}$ is the brownian velocity of the gas particles where $\Boltzmann$ is the Boltzmann constant, $d_{i}$ is the mean particle diameter in the $i^{\text{th}}$ section, $\alpha_{\textsc{t}}$ is the thermal accomodation factor representing the fraction of the gas molecules that leave the surface in equilibrium with the surface, the remaining fraction $1-\alpha_{\textsc{t}}$ being specularly reflected: this constant is usually taken equal to $\alpha_{\textsc{t}}=0.9$ \cite{waldmann_thermophoresis_1966,friedlander_smoke_2000}.

The particle thermophoresis velocity is taken from Waldmann \cite{waldmann_thermophoresis_1966} as $\boldsymbol{v}_\text{th} = - C_{\text{th}} \, \nu \, \boldsymbol{\nabla} \ln{T}$, where $C_{\text{th}} = 3/4 (1 + \pi \alpha_{\textsc{t}}/8)^{-1} \approx 0.554 $, and $\nu$ is the gas kinematic viscosity.

In the non-conservative formulation, which is often used when the nanoparticle concentration is low \cite{zhao_measurement_2003,wang_experiments_1996,attili_formation_2014,jocher_impact_2017,rodrigues_unsteady_2017}, in general only the gaseous species diffusion velocities are corrected, namely \cite{wang_modeling_2005}:
\begin{gather}
\boldsymbol{\Vel} _{k} = - D_{k} \, \boldsymbol{\nabla} \ln{X_{k}} + \boldsymbol{u}_{\text{cor}}^{\text{g}}, \label{eq:TwoMixtDiscDiffFluxGas} \\
\boldsymbol{\Vel} _{i}= \boldsymbol{v}_\text{th} - D_{i} \boldsymbol{\nabla} \ln{Y_{i}}, \label{eq:TwoMixtDiscDiffFluxPart}
\end{gather}
where $\boldsymbol{u}_{\text{cor}}^{\text{g}}$ is a gaseous correction velocity, taken such that the following constraint is satisfied:
\begin{equation}
 \sum_{k=1}^{N_{\text{g}}} Y_{k} \boldsymbol{\Vel} _{k} = 0. \label{eq:TwoMixtDiscDiffFluxGasCons}
\end{equation}

It is clear that this set of equations is not conservative, potentially leading to large errors or numerical instabilities when used for highly concentrated aerosols. In the following, this 'non-mass-conserving' formulation will be compared to the conservative formulation, Eqs. \eqref{eq:OneMixtDiscDiffFluxGasCons} and \eqref{eq:OneMixtDiscDiffFluxPartCons}.

\subsection{Enthalpy}
\label{subsec:enthalpy}

As well, two formulations may be used for the mixture enthalpy, a conservative and a non-conservative one.

In the conservative formulation, the mixture specific enthalpy is given by
\begin{equation}
h = \sum_{k=1}^{N_{\text{g}}} Y_{k} h_{k} + \sum_{i=1}^{N_{\text{s}}} Y_{i} h_{i},
\label{eq:OneMixtDiscEnthalpy}
\end{equation}
where $h_{k}$ is the $k^{\text{th}}$ species specific enthalpy and $h_{i}$ is the average specific enthalpy of particles in the $i^{\text{th}}$ section. The heat flux reads:
\begin{equation}
\boldsymbol{Q} = - \lambda \boldsymbol{\nabla} \T + \sum_{k=1}^{N_{\text{g}}} \rho Y_{k} h_{k} \boldsymbol{\Vel} _{k}+ \sum_{i=1}^{N_{\text{s}}} \rho Y_{i} h_{i} \boldsymbol{\Vel} _{i}, \label{eq:OneMixtDiscHeatFlux}
\end{equation}
where $\lambda$ is the global thermal conductivity of the -- gas and particles -- mixture. Here, we classically neglect the Dufour effect, although it could be accounted for straightforwardly. Note that the mixture specific enthalpy -- Eq. \eqref{eq:OneMixtDiscEnthalpy} -- can be breaken down into gas and particle contributions:
\begin{equation}
h = \tilde{h}_{\text{g}} + \tilde{h}_{\text{p}},
\label{eq:OneMixtContEnthalpyDiss}
\end{equation}
where $\tilde{h}_{\text{g}}$ and $\tilde{h}_{\text{p}}$ are the respective contributions of the gas and particle phases to the mixture specific enthalpy, which read:
\begin{gather}
\tilde{h}_{\text{g}} = Y_{\text{g}} h_{\text{g}} = \sum_{k=1}^{N_{\text{g}}} Y_{k} h_{k}, \\
\tilde{h}_{\text{p}} = Y_{\text{p}} h_{\text{p}} = \sum_{i=1}^{N_{\text{s}}} Y_{i} h_{i}, \\
\end{gather}
where $h_{\text{g}}$ and $h_{\text{p}}$ are the gas and particle specific enthapies, respectively.

In the non-conservative formulation, the mixture specific enthalpy is computed as \cite{wang_modeling_2005}:
\begin{equation}
h=\sum_{k=1}^{N_{\text{g}}} Y_{k}h_{k},
\label{eq:OneMixtContEnthalpyNonCons}
\end{equation}
and the contribution of the nanoparticles $\sum_{i=1}^{N_{\text{s}}} \rho Y_{i} h_{i} \boldsymbol{\Vel} _{i}$ to the heat flux is neglected in Eq. \eqref{eq:OneMixtDiscHeatFlux}. In the present work, this 'non-energy-conserving' formulation will be compared to the conservative formulation, Eqs. \eqref{eq:OneMixtDiscEnthalpy} and \eqref{eq:OneMixtDiscHeatFlux}. The different formulations envisioned are synthesized in Tab. \ref{tab:Synthetic} 

\begin{table}[H]
\begin{small}
\begin{center}
\begin{tabular}{c|c|c}

        & \textbf{Conservative} & \textbf{Non-conservative} \\
	\hline
	\textbf{Diffusion} & $\displaystyle \sum_{k=1}^{N_{\text{g}}} \rho Y_{k} \boldsymbol{\Vel} _{k}+ \sum_{i=1}^{N_{\text{s}}} \rho Y_{i} \boldsymbol{\Vel} _{i}= 0$ & $\displaystyle \sum_{k=1}^{N_{\text{g}}} \rho Y_{k} \boldsymbol{\Vel} _{k} = 0$ \\
	\hline
	\textbf{Enthalpy} & $\displaystyle h = \sum_{k=1}^{N_{\text{g}}} Y_{k} h_{k} + \sum_{i=1}^{N_{\text{s}}} Y_{i} h_{i}$ & $\displaystyle h = \sum_{k=1}^{N_{\text{g}}} Y_{k} h_{k}$ \\
	\hline

\end{tabular}
\end{center}
\end{small}
\caption{Summary of the different formulations investigated in the present study.}
\label{tab:Synthetic}
\end{table}

It is well known that, in the presence of nanoparticles or soots, radiation can strongly modify the flame structure, e.g. due to radiative heat loss induced by nanoparticles emission. Yet to remain as simple as possible, we chose not to account for radiation. Indeed, although radiation will strongly influence the flame and nanoparticles profiles, it should not change qualitatively the impact of mass and enthalpy conservation.

\section{Titania synthesis in flames}
\label{sec:applications}

In order to demonstrate the importance of the conservative formulation, the synthesis of titania (TiO$_{2}$) nanoparticles is considered here as a case illustration. Any other kinds of fine particles that can be produced in flames could be addressed, yet titania particles present a relatively high conversion yield so it is natural to consider this type of particles. For the present purpose, the nucleation kinetics is well described by a one-step nucleation model, which is such that nucleation is a complete and fast reaction. Such a model represents well the rapidity of the nucleation process, without fine details on the real nucleation pathways.

Physical processes involved -- i.e. nucleation, coagulation, and surface growth -- are described in this section. No sintering is considered since the shape of the nanoparticles produced is not relevant for the purpose of the present study. The source term in Eq. \eqref{eq:OneMixtDiscParticleCons} thus reads:
\begin{equation}
\dot{Q}_{i} = \dot{Q}_{\text{nu},i} + \dot{Q}_{\text{coag},i} + \dot{Q}_{\text{sg},i}.
\label{eq:sumparticlesrct}
\end{equation}

\subsection{Nucleation model}

One of the main precursors used in industrial processes for flame synthesis of TiO$_{2}$ nanoparticles is titanium tetrachloride (TiCl$_{4}$). Pratsinis et al. \cite{pratsinis_kinetics_1990} first described the oxidation of TiCl$_{4}$ vapor between 700 and 1000{\degres}C as a one-step chemical reaction:
\begin{equation}
\text{TiCl}_{4}\text{(g)} + \text{O}_{2}\text{(g)} \rightarrow \text{TiO}_{2}\text{(s)} + 2 \text{Cl}_{2}\text{(g)}
\label{OneStep}
\end{equation}
The rate is first-order with respect to TiCl$_{4}$, and nearly zeroth-order for O$_{2}$ up to a 10-fold oxygen excess. Accordingly, the titania nanoparticle nucleation rate can be written in the following form \cite{pratsinis_kinetics_1990,pratsinis_competition_1998}:
\begin{equation}
\dot{Q}_{\text{nu},i} = \delta_{i1} \, \ChemK_{1\text{step}} \n_{\text{TiCl}_{4}} \, v_{\text{TiO}_{2}},
\label{OneStepRate}
\end{equation}
where $\delta_{i1}$ equals $1$ if $i=1$ and $0$ otherwise, $\n_{\text{TiCl}_{4}}$ is the TiCl$_{4}$ number density, $v_{\text{TiO}_{2}} = \MoleMass_{\text{TiO}_{2}}/(\Avogadro \rho_{\text{s}})$ is the volume of a monomer, and $\ChemK_{1\text{step}}$ is the one-step rate constant, which reads:
\begin{equation}
\ChemK_{1\text{step}}=A_{1\text{step}} \exp{(-\ActivationEnergy_{1\text{step}}/(\RU \T))}
\label{OneStepConst}
\end{equation}
with $A_{1\text{step}} = 8.26 \cdot 10^{4}$ s$^{-1}$ and $\ActivationEnergy_{1\text{step}} = 88.8$~kJ.mol$^{-1}$ the activation energy. This rate has been the basis of many numerical studies of TiO$_{2}$ nanoparticle formation \cite{xiong_formation_1993,pratsinis_competition_1998,spicer_titania_2002,tsantilis_narrowing_2004,wang_modeling_2005,morgan_new_2006,heine_agglomerate_2007}.

As the purpose of this paper is essentially to demonstrate the importance of using a conservative set of equations, the one-step nucleation scheme is adopted here in conjunction with the classical GRI-Mech 3.0 \cite{smith_gri-mech_nodate} for the oxidation of methane. This kinetic mechanism has indeed been extensively validated against experimental data \cite{xiong_formation_1993,pratsinis_competition_1998,spicer_titania_2002,tsantilis_narrowing_2004,morgan_new_2006,heine_agglomerate_2007,xu_cfd-population_2017}, although most of those validations focused on the TiCl$_{4}$ consumption rate rather than on the nanoparticle production yield. The global scheme adopted here is expected to be sufficient to recover the correct order of magnitude of conversion rate, although nanoparticles nucleation, coagulation, or surface growth characteristic timescales may be slightly misestimated. Following the recommandations of Mehta et al. \cite{mehta_multiscale_2010}, we consider the nucleated particles to contain five Ti atoms. In other words, the smallest volume of the first section is taken equal to $v_{1}^{\text{min}} = 5 \, v_{\text{TiO}_{2}}$. The thermodynamic and transport data for TiCl$_{4}$, Cl$_{2}$ and TiO$_{2}$ are taken from \cite{mehta_reduced_2015}. The density of titania nanoparticles is assumed constant, equal to $\rho_{\text{s}} = 4000$~kg.m$^{-3}$ \cite{pratsinis_competition_1998}. Here, for the sake of simplicity, we consider that the thermal conductivity $\lambda$ is equal to that of the gas phase alone.

\subsection{Coagulation}
The global coagulation source term is expressed according to the Smoluchowski's expression \cite{gelbard_sectional_1980}:
\begin{equation}
\dot{Q}_{\text{coag},i} = \bigg( \sum_{1 \leq j \leq k}^{i} \dot{N}_{coag}^{j,k \rightarrow i} - \sum_{j=1}^{N_{\text{s}}} \dot{N}_{ij}^{\text{out}} \bigg) \frac{Q_{i} }{N_{i} },
\end{equation}
where $\dot{N}_{\text{coag}}^{j,k\rightarrow i}$ is the number of particles received by the $i^{\text{th}}$ section due to collisions of particles from the $j^{\text{th}}$ and $k^{\text{th}}$ sections per unit time and $\dot{N}_{ij}^{\text{out}}$ is the number of particles leaving the $i^{\text{th}}$ section upon collision with particles of the $j^{\text{th}}$ section per unit time.

\par
The collision frequency $\beta_{i,j}$ between a particle of the $i^{\text{th}}$ section and a particle of the $j^{\text{th}}$ section is evaluated at the average volumes $v^{\text{mean}}_{i}$ and $v^{\text{mean}}_{j}$. Here, a transition regime between the free molecular regime (superscript $^{\text{fm}}$) and the continuum regime (superscript $^{\text{c}}$) has been chosen for the description of collisions, so that $\beta_{i,j}$ is expressed as:
\begin{equation}
\beta_{i,j} = \frac{\beta_{i,j}^{\text{fm}}\beta_{i,j}^{\text{c}}}{\beta_{i,j}^{\text{fm}}+\beta_{i,j}^{\text{c}}} \approx \text{min}(\beta_{i,j}^{\text{fm}},\beta_{i,j}^{\text{c}})
\end{equation}
with:
\begin{equation}
\begin{aligned}
&\beta_{i,j}^{\text{fm}} = && \frac{1}{2} \, \epsilon_{\text{coag}} \, \left(\frac{2\pi \Boltzmann T}{\rho_{\text{s}}}\right)^{1/2} \sqrt{\frac{1}{v_{i}^{\text{mean}}} + \frac{1}{v_{j}^{\text{mean}}}}(d_{c,i}+d_{c,j})^{2} \\
&\beta_{i,j}^{\text{c}} = &&\frac{2 \Boltzmann T}{3\mu}\left(d_{c,i}+d_{c,j}\right)\left(\frac{\text{C}_{i}}{d_{c,i}} + \frac{\text{C}_{j}}{d_{c,j}}\right)
\end{aligned}
\label{eq:betafmbetac}
\end{equation}
where $ \epsilon_{\text{coag}} = 2.2 $ is an amplification factor due to Van der Waals interactions \cite{marchal_modelisation_2008,rodrigues_modelisation_2018} and $\mu $ is the gas dynamic viscosity given by Sutherland's formula \cite{sutherland_viscosity_1893} $\mu = C_{1} \T^{3/2}/(\T+C_{2})$. The coefficients $C_{1}$ and $C_{2}$ are the Sutherland coefficients, and $\text{C}_{j}$ is the Cunningham corrective coefficient for a particle of the $j^{\text{th}}$ section \cite{cunningham_velocity_1910,pratsinis_simultaneous_1988}:
\begin{equation}
\text{C}_{j} = 1 + 1.257 \, \text{Kn}_{j} = 1 + 1.257 \, \frac{2l_{\text{gas}}}{d_{c,j}}
\end{equation}
where $\text{Kn}_{j}$ is the Knudsen number. $d_{c,j}$ is the collisional diameter of a particle of the $j^{\text{th}}$ section, considered constant and evaluated as a function of $n_{j}$, $d_{j}$ and the fractal dimension $D_{f}$ of particles is defined from the relation $d_{c,j} = d_{j}n_{j}^{1/D_{f}}$. As we consider here spherical particles, the fractal dimension $D_{f}$ is equal to $1$ and $d_{c,j} = d_{j}$. Finally, $l_{\text{gas}} = \Boltzmann T / (\sqrt{2}\pi d^{2}_{\text{gas}}\p)$ is the mean free path of the gaseous phase,
where $\Boltzmann$ is the Boltzmann constant, $d_{\text{gas}}=0.2~\text{nm}$ is the diameter of a typical gas particle and $\p$ is the pressure.

\subsection{Surface growth}
\label{subsec:sgrowth}

With the one-step kinetics employed here, and under the conditions considered, the contribution of surface growth \cite{ghoshtagore_mechanism_1970,shirley_theoretical_2011} is marginal as nucleation is fast and consumes most of the TiCl$_{4}$ rapidly: $\dot{Q}_{\text{sg},i}=0$.

\section{Validation}
\label{subsec:validation}

The model has been validated using the experimental results of Nakaso et al. \cite{nakaso_size_2001} obtained on a furnace reactor. In this experiment, TiCl4 is injected in an O2/N2 mixture flowing inside a heated tube. Here, the configuration is described as 1-D premixed case, by imposing the experimental temperature profile provided in Figure 4 in \cite{nakaso_size_2001} for a furnace temperature of $T_{f}$=1200~K. The N$_{2}$ and O$_{2}$ mole fractions are 75~\% and 25~\%, respectivel. The TiCl$_{4}$ mole fraction is $2 \cdot 10^{-5}$. The flow rate is $2.74 \cdot 10^{-2}$~g.cm$^{-2}$.s$^{-1}$.

\begin{figure}[H]
\begin{minipage}[t]{.5\textwidth}
 \begin{center}
\subfloat[Nanoparticle agglomerates size distribution functions from calculations by Nakaso et al. \cite{nakaso_size_2001}.]{\includegraphics[height = 85mm, clip = true]{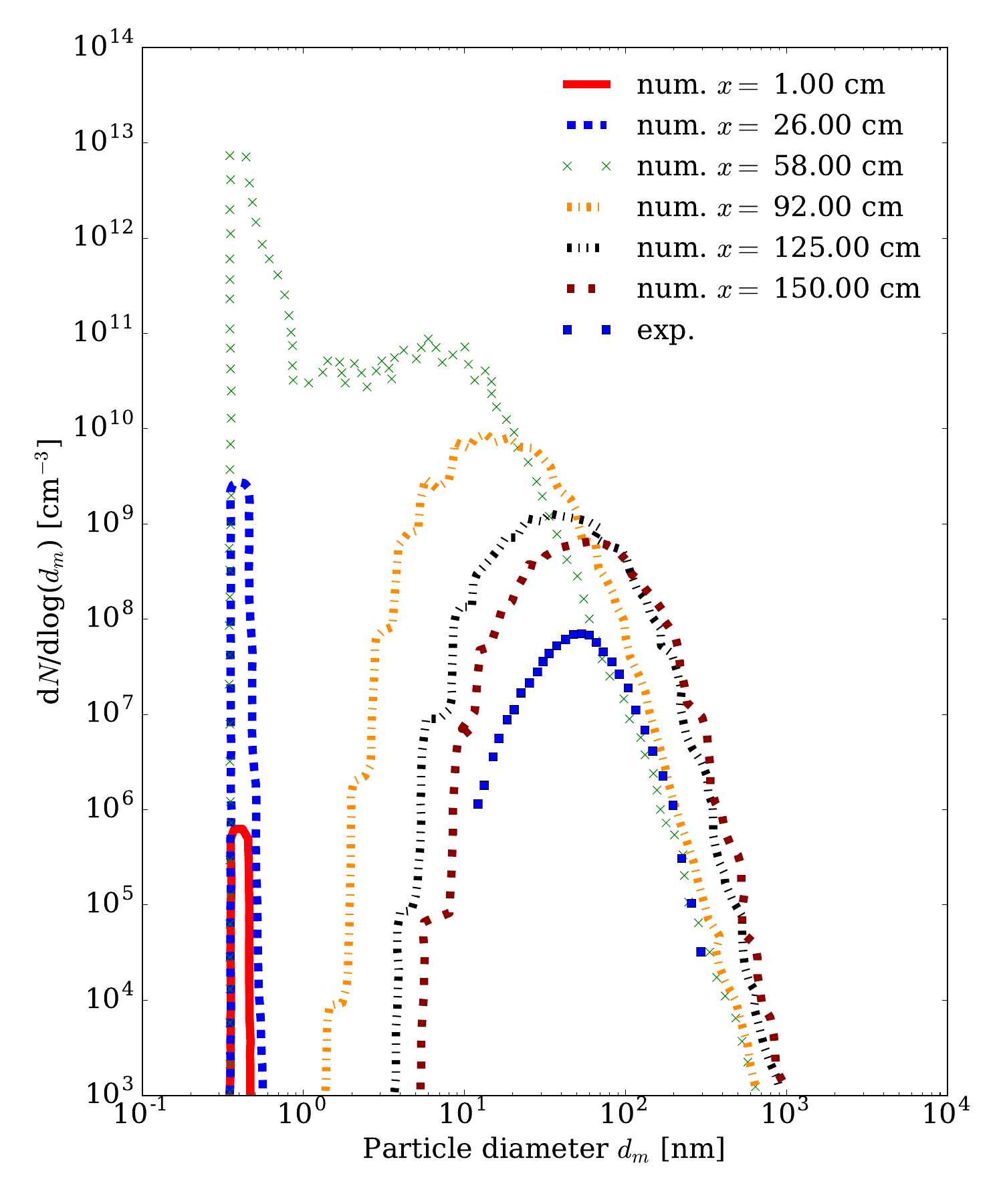}\label{fig:psdf_nakaso}}
\end{center}
\end{minipage}%
\begin{minipage}[t]{.5\textwidth}
\begin{center}
\subfloat[Nanoparticle agglomerates size distribution functions calculated using the present model.]{\includegraphics[height = 85mm, clip = true]{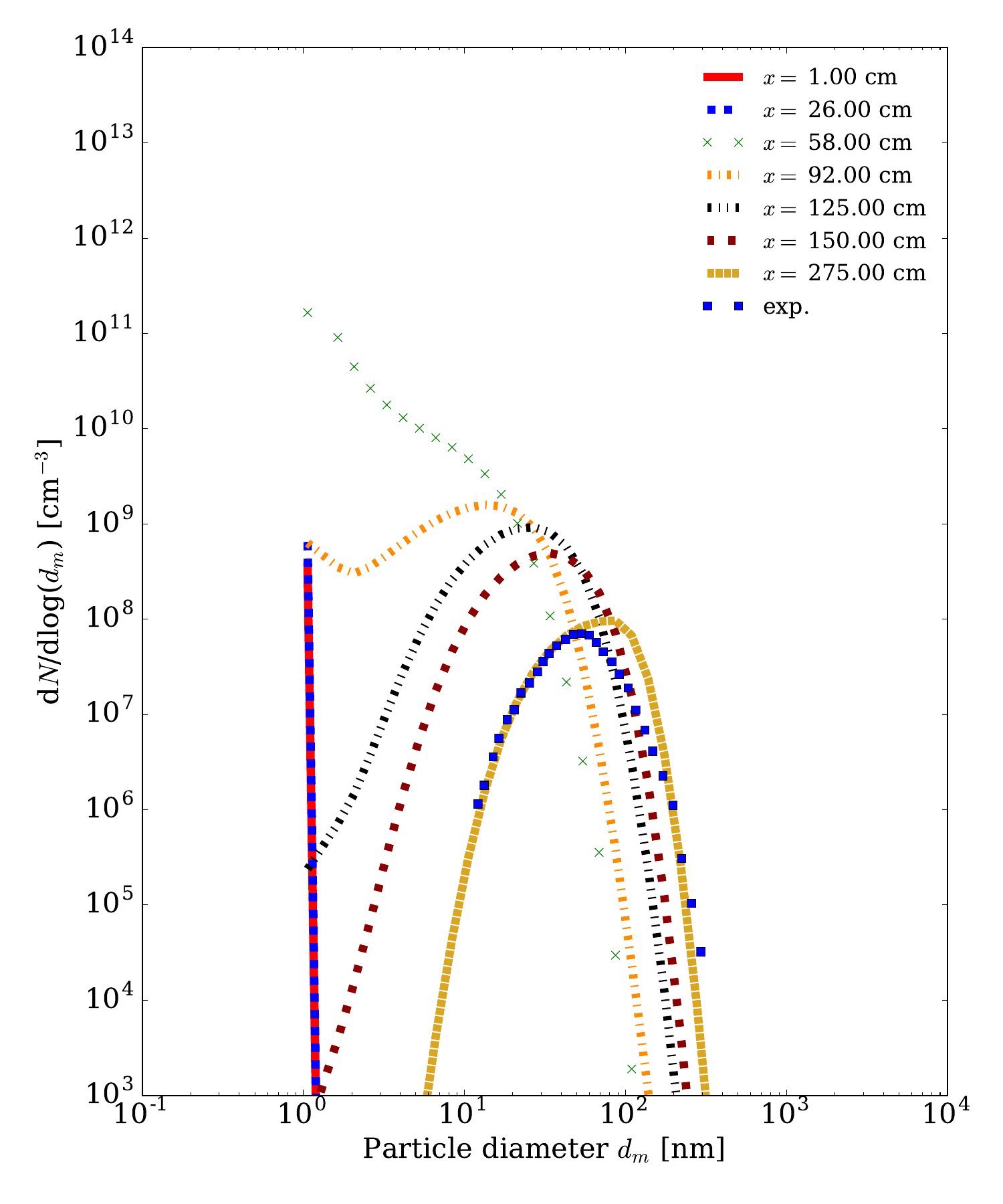}\label{fig:psdf}}
\end{center}
\end{minipage}
\caption{Comparison of simulations with experimental measurements on the configuration of Nakaso et al. for $T_{f} = 1200$~K \cite{nakaso_size_2001}.}
\label{fig:validation results}
\end{figure}

Figure \ref{fig:psdf_nakaso} shows the particle size distribution (PSD) functions obtained by Nakaso et al. using a trivariate (diameter/volume/surface) sectional model at different positions together with the experimental PSD which is measured downstream the furnace exit.  Figure \ref{fig:psdf} shows the PSDs obtained with the present model. It can be concluded that the description retained in the present work reproduces the literature results. It is worth noting that the original experiment of Nakaso et al. uses a 1.5m-long tube but the nanoparticle detection systems are located downstream the exit of the tube, which may explain the discrepancy between the measured distribution and those calculated at $x=1.5$~m by Nakaso et al. \cite{nakaso_size_2001} as well as by the present results. For that reason, we have also plotted in Figure \ref{fig:psdf} the calculated distribution at $x = 2.75$~m, which appears to be very close to the experimental distribution.

\section{Results in 1D premixed configuration}
\label{sec:1dp}

In this section and the following one, we evaluate the importance of the modelling -- conservative v.s. non-conservative -- on simple idealized laminar flames. Such flames are far from typical industrial configurations for titania nanoparticle productions which generally rely on highly turbulent flames in order ro enhance mixing and reactivity. The objective here is not to give some insights on the experimental process, but rather to evaluate, in a modelling perspective, the impact of the model chosen. In this respect, it is desirable to consider first simple flame configurations to reduce the physical complexity and thus to better evaluate the impact of the model. Besides, many turbulent models rely on such laminar flames models which are supposed to describe well the local flame structure. Therefore, studying premixed and counterflow flames is a necessary step towards the modelling of more complex turbulent flames.

First, in this section, we measure the impact of conservation when high concentrations of TiO$_{2}$ are encountered in a reactive premixed flow. For this, one-dimensional (1D) premixed CH$_{4}$/TiCl$_{4}$/O$_{2}$/N$_{2}$ flames are calculated using the 1D premixed model in the in-house code Regath \cite{rodrigues_unsteady_2017, rodrigues_modelisation_2018}, using both conservative and non-conservative formulations for enthalpy. In this configuration, the contribution of particles diffusion is expected to be negligible so that the mass-conservative formulation is retained for all computations of this section. In the configuration studied here, the low Mach number approximation applies and the momentum equation is not needed. The CH$_{4}$/O$_{2}$ mixture is at stoichiometric conditions with a CH$_{4}$ inlet mass fraction of 5.5\,\%.

The working pressure is the atmospheric pressure, and the mixture is pre-heated at 500\,K so that the TiCl$_{4}$ is fully vaporized. TiCl$_{4}$ inlet mass fraction is equal to $5 \, \%$, the O$_{2}$ inlet mass fraction is 22.0\,\% and the N$_{2}$ mass fraction is 67.5\,\%. With such a high nanoparticle mass fraction, it is expected that the phase change has some impact on the gas-phase enthalpy, and possibly on the flame structure.

\begin{figure}[H]
\begin{minipage}[t]{.5\textwidth}
 \begin{center}
\subfloat[Main species mass fractions and temperature.]{\includegraphics[height = 45mm, clip = true]{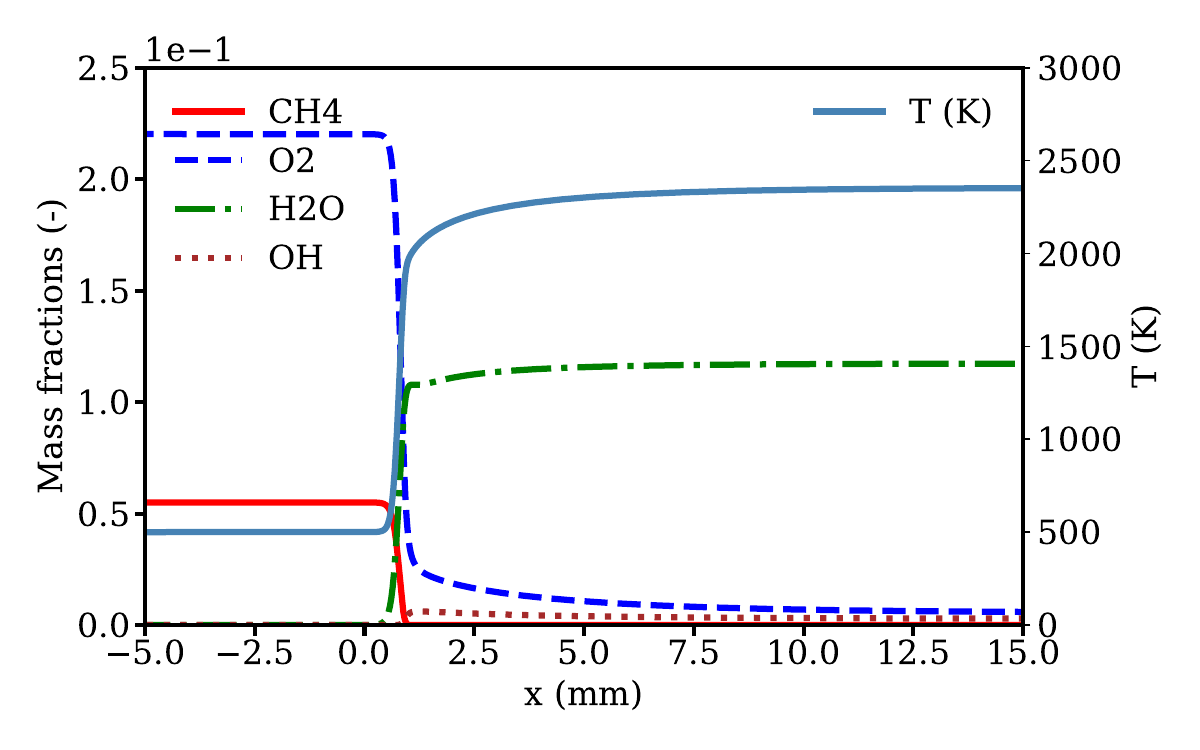}\label{fig:1dp_grionestep_flame_phi1_ticl4_5e-2}}
\end{center}
\end{minipage}%
\begin{minipage}[t]{.5\textwidth}
\begin{center}
\subfloat[Titania species mole fractions.]{\includegraphics[height = 45mm, clip = true]{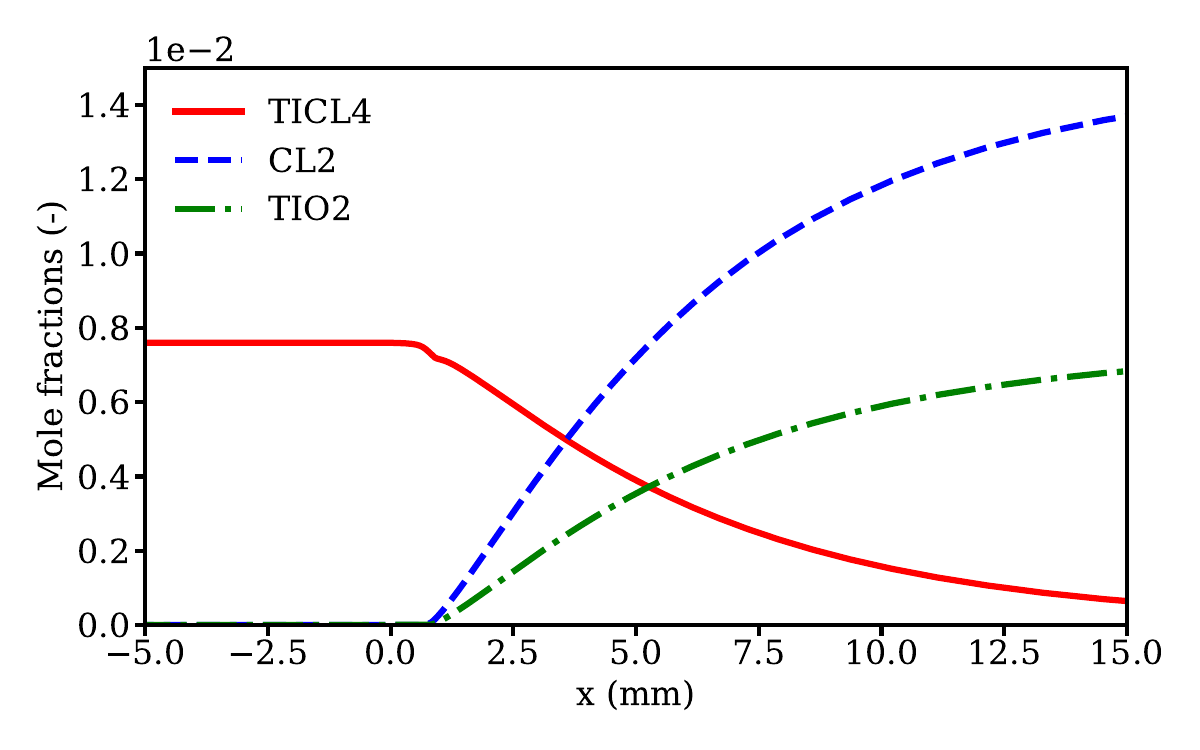}\label{fig:1dp_grionestep_xtitania_phi1_ticl4_5e-2}}
\end{center}
\end{minipage}
\caption{Main species and temperature profiles for the 1D premixed CH$_{4}$/TiCl$_{4}$/O$_{2}$/N$_{2}$ flame at stoichiometric conditions. The inlet temperature is 500~K. Injected TiCl$_{4}$ mass fraction $Y_{\text{TiCl}_{4}}^{\text{inj}}=5 \cdot 10^{-2}$. Results are obtained using the conservative model.} \label{fig:1dp_grionestep_phi1_ticl4_5e-2}
\end{figure}

Figs. \ref{fig:1dp_grionestep_flame_phi1_ticl4_5e-2} and \ref{fig:1dp_grionestep_xtitania_phi1_ticl4_5e-2} show respectively the main combustion species mass fraction and temperature profiles, and the TiCl$_{4}$, Cl$_{2}$ and TiO$_{2}$ mole fraction profiles in the 1D premixed flame. Although the mass fractions are generally the variables of interest, as they are the transported variables, in the present case the Ti-containing species mole fractions are plotted rather than the mass fractions, as the number of Ti atoms is conserved so that the conversion yield can be visualized more easily in terms of mole fractions. As it can be seen in Fig. \ref{fig:1dp_grionestep_xtitania_phi1_ticl4_5e-2}, the conversion of TiCl$_{4}$ into TiO$_{2}$ is very efficient, almost equal to $100 \%$. The TiO$_{2}$ one-step reaction is relatively fast, although the reaction front is much less stiff than the combustion front depicted in Fig. \ref{fig:1dp_grionestep_flame_phi1_ticl4_5e-2}.

\begin{figure}[H]
\begin{center}
\includegraphics[height = 60mm, clip = true]{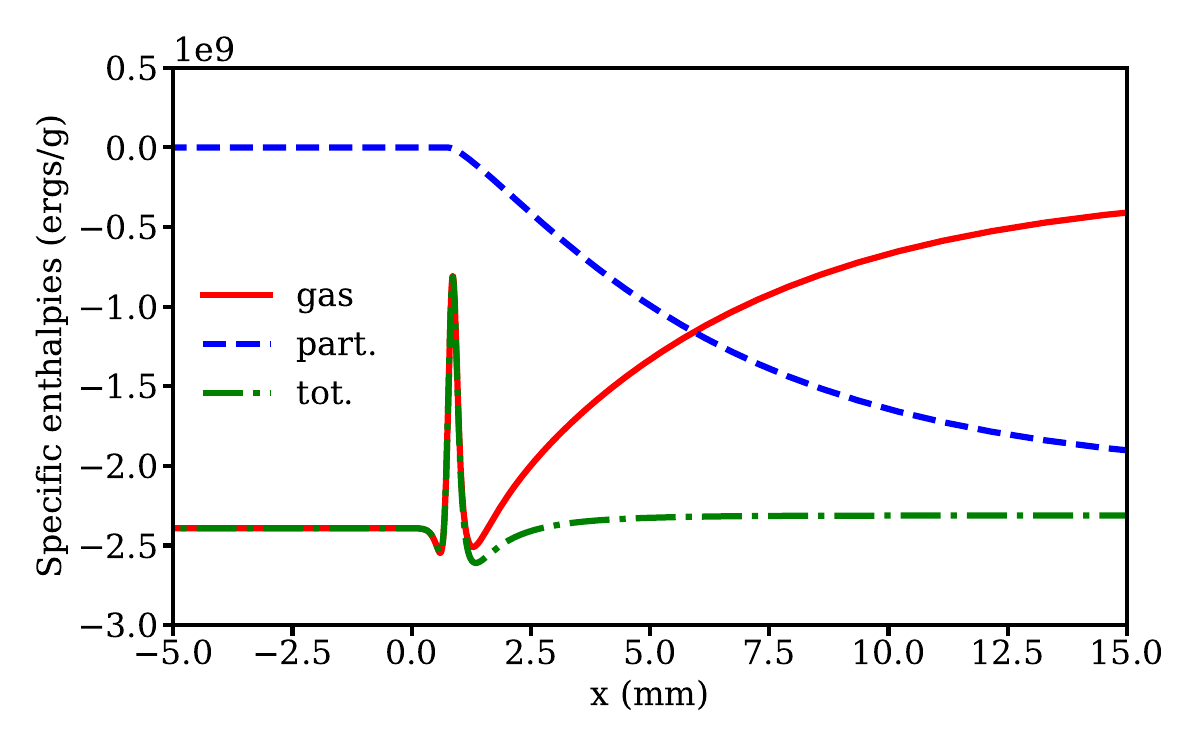}
\caption{Enthalpies of the gas phase $\tilde{h}_{\text{g}}$, the solid particle phase $\tilde{h}_{\text{p}}$, and the mixture $h=\tilde{h}_{\text{g}}+\tilde{h}_{\text{p}}$. Same conditions as in Fig. \ref{fig:1dp_grionestep_phi1_ticl4_5e-2}.}
\label{fig:1dp_grionestep_hk_phi1_ticl4_5e-2}
\end{center}
\end{figure}

In Fig. \ref{fig:1dp_grionestep_hk_phi1_ticl4_5e-2} the enthalpies of the respective phases -- gas/particle -- and the total enthalpy of the mixture are plotted. The gas enthalpy $\tilde{h}_{\text{g}}$ increases significantly as the gaseous TiCl$_{4}$ is converted into solid TiO$_{2}$. In the burnt gases, most of the mixture enthalpy $h$ comes from the particle enthalpy $\tilde{h}_{\text{p}}$, even though the injected mass fraction of TiCl$_{4}$ is only $5\,\%$. This is due to the relatively large absolute value of the enthalpy of TiO$_{2}$ \cite{west_first-principles_2007}. Note that the total enthalpy is conserved in the 1D premixed flame with the conservative model, proving that the model is indeed conservative.

\begin{figure}[H]
\begin{center}
\includegraphics[height = 60mm, clip = true]{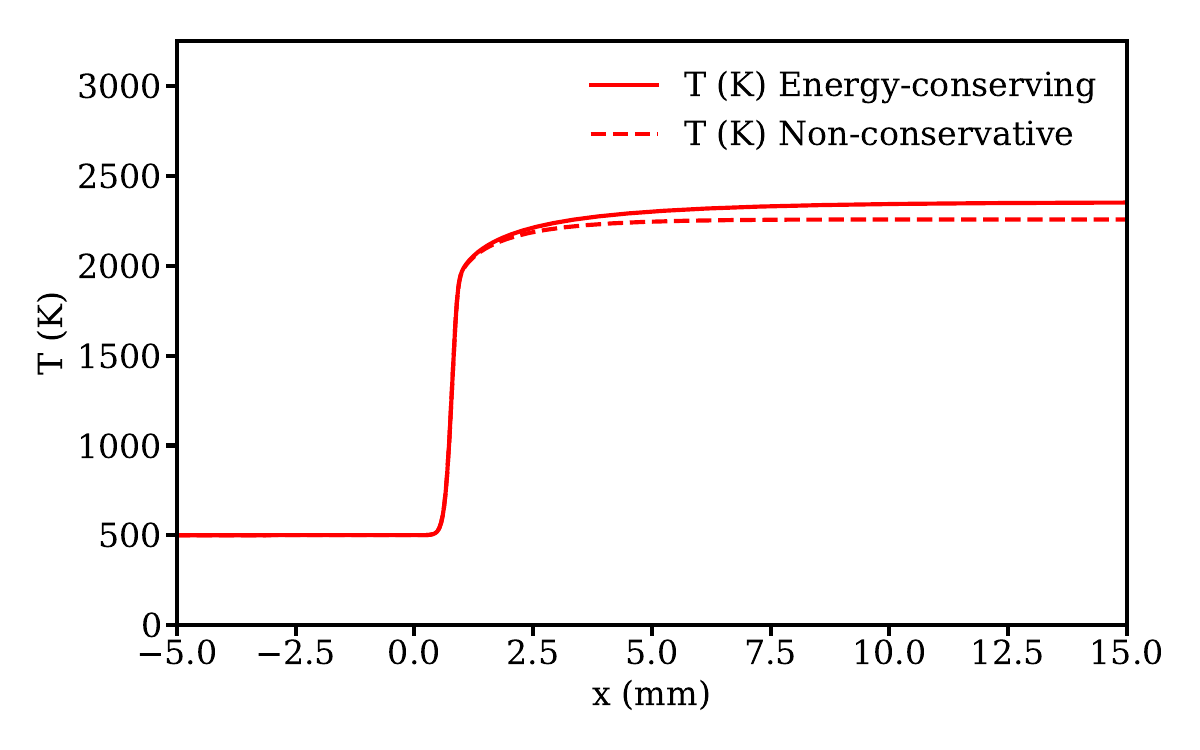}
\caption{Comparison of the temperature profiles obtained with the energy-conserving model and the non-energy-conserving one. Same conditions as in Fig. \ref{fig:1dp_grionestep_phi1_ticl4_5e-2}.}
\label{fig:1dp_grionestep_temperature_phi1_ticl4_5e-2_Comp_Cons_nonCons}
\end{center}
\end{figure}

If one neglects the enthalpy of the particle phase $\tilde{h}_{\text{p}}$ in expression Eq. \eqref{eq:OneMixtContEnthalpyDiss}, then the set of equations is non-conservative in essence and may yield non-physical results. Indeed, exchanges of enthalpy occur between the gas and the particle phases, and thus neglecting $\tilde{h}_{\text{p}}$ yields for instance an erroneous temperature in the burnt gases. In Fig. \ref{fig:1dp_grionestep_temperature_phi1_ticl4_5e-2_Comp_Cons_nonCons} the temperature profile obtained with the energy-conserving model -- Eqs. \eqref{eq:OneMixtDiscEnthalpy} and \eqref{eq:OneMixtDiscHeatFlux} -- is compared with the one obtained using the non-energy-conserving formulation -- Eq. \eqref{eq:OneMixtContEnthalpyNonCons}. The mass-conserving model is kept in both calculations. As expected the flame structure is impacted. For only $5 \%$ of TiCl$_{4}$, the adiabatic temperature is reduced by $95 \,$K when the non-conservative model is used compared to the conservative formulation. The effect on flame temperature might appear modest for these conditions, as the adiabatic temperature is only modified by 7 \%. However, for higher injection rates, typical of industrial conditions, the effect is expected to be even higher.

\begin{figure}[H]
\begin{center}
\includegraphics[height = 60mm, clip = true]{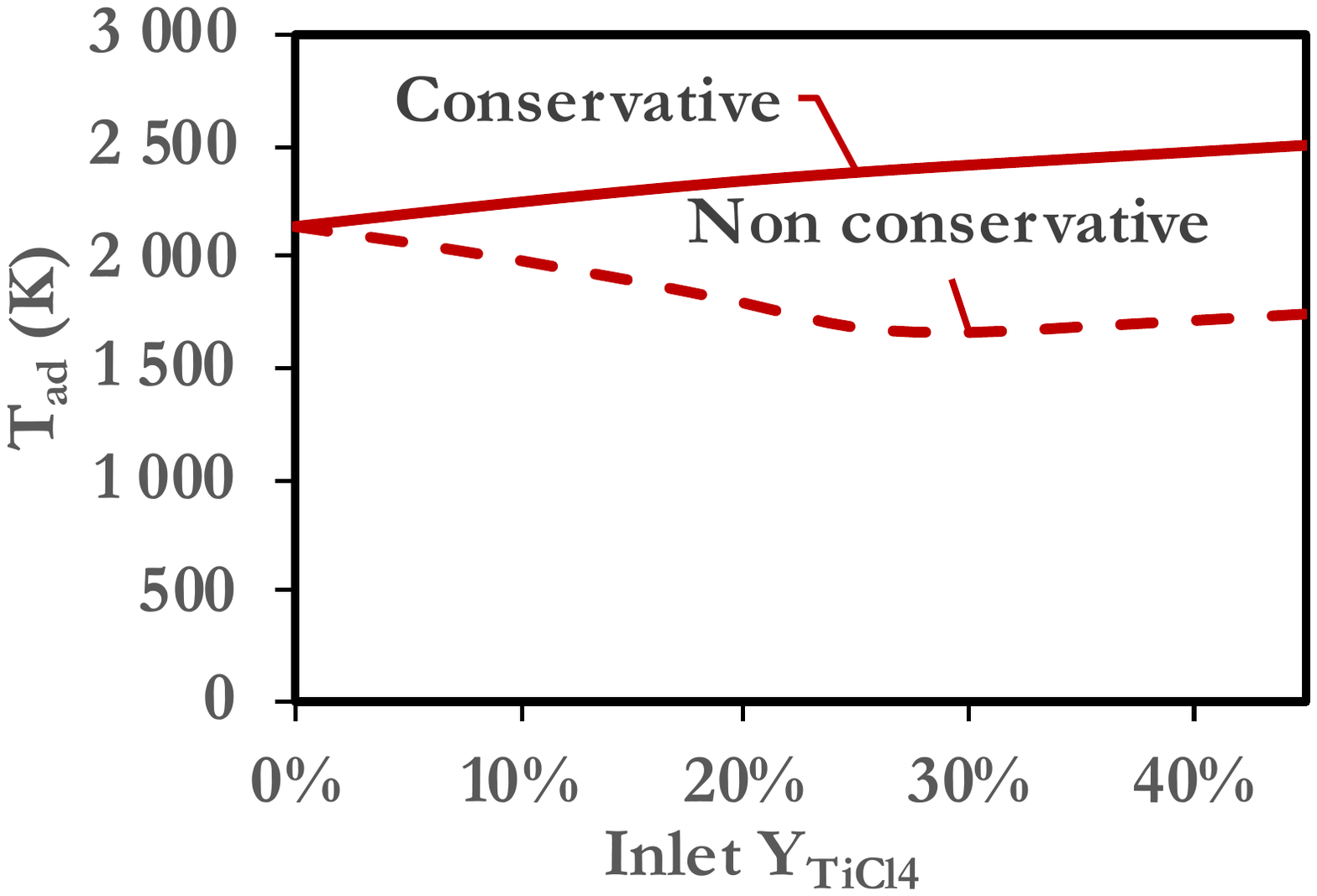}
\caption{Comparison of the burnt gas adiabatic temperatures obtained with the energy-conserving model (continuous line) and the non-energy-conserving one (dashed line), for varying values of inlet TiCl$_{4}$ mass fraction. The fresh gas composition corresponds to $\phi = 0.8$.}
\label{fig:1dp_grionestep_phi08_temperature}
\end{center}
\end{figure}

\begin{figure}[H]
\begin{minipage}[t]{.5\textwidth}
 \begin{center}
\subfloat[Main titania-related species mass fractions.]{\includegraphics[height = 50mm, clip = true]{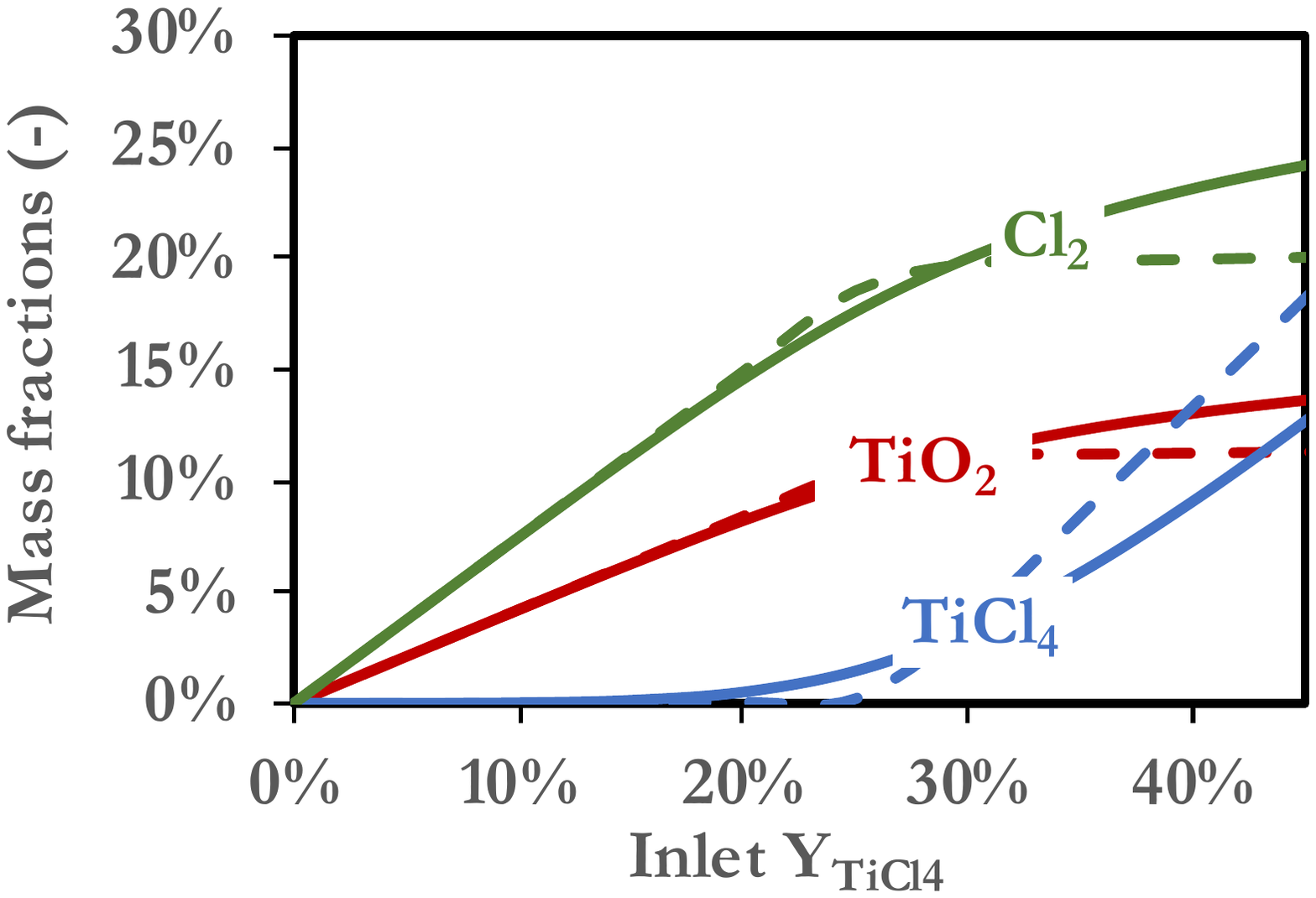}\label{fig:1dp_grionestep_phi08_ytitania}}
\end{center}
\end{minipage}%
\begin{minipage}[t]{.5\textwidth}
\begin{center}
\subfloat[Combustion products mass fractions.]{\includegraphics[height = 50mm, clip = true]{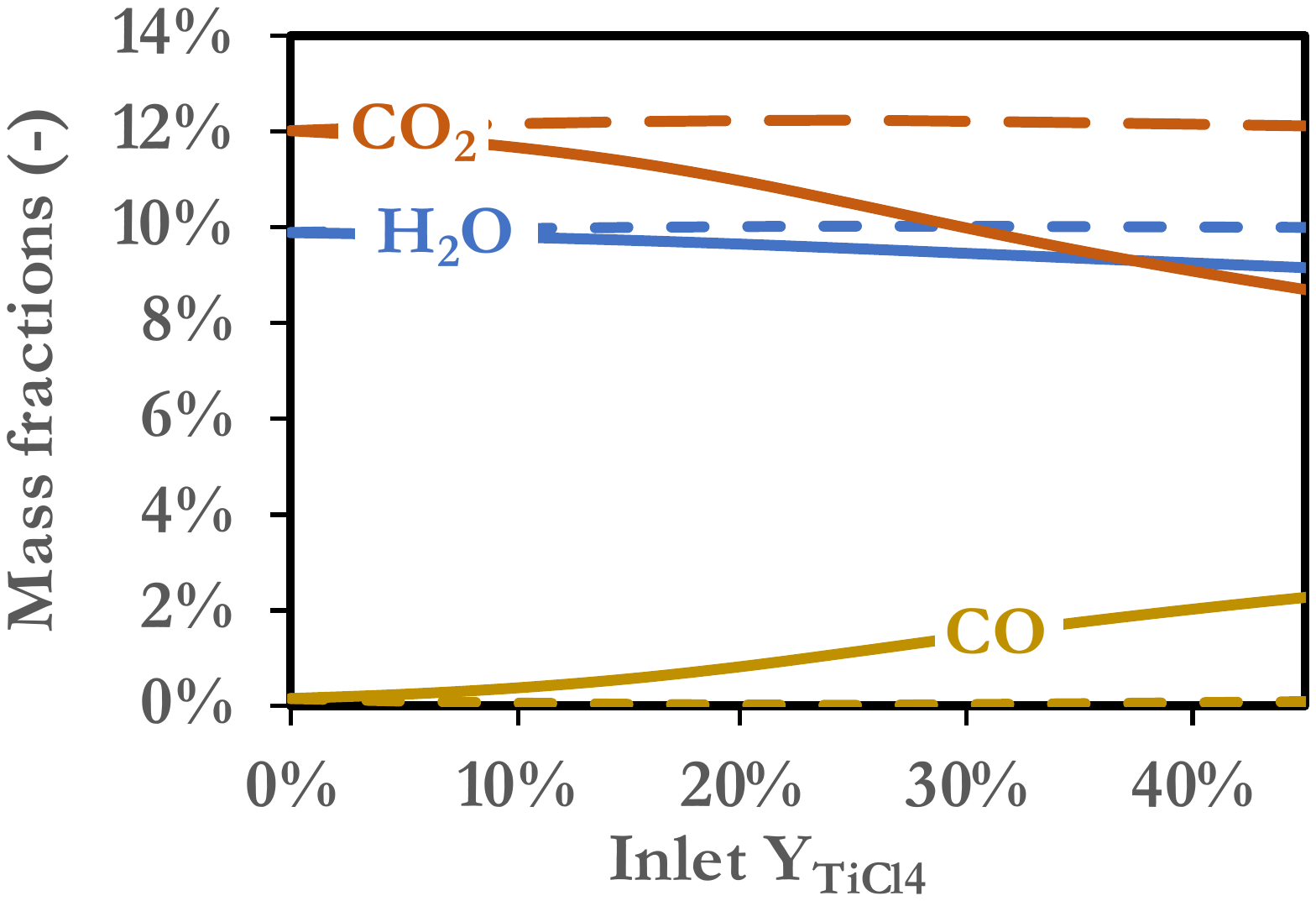}\label{fig:1dp_grionestep_phi08_yflame}}
\end{center}
\end{minipage}
\caption{Comparison of the burnt gas mass fractions obtained with the energy-conserving model (continuous lines) and the non-energy-conserving one (dashed lines), for varying values of inlet TiCl$_{4}$ mass fraction. The fresh gas composition corresponds to $\phi = 0.8$.}
\label{fig:1dp_grionestep_phi08_yk}
\end{figure}

Unfortunately, this cannot be verified by performing 1D premixed flames with higher TiCl$_{4}$ concentration, since when increasing the TiCl$_{4}$ inlet concentration, the non-conservative 1D calculations do not converge anymore because this formulation is numerically unstable. Yet thermodynamic equilibrium calculations have been carried out as an alternative to quantify the effect of a non-conservative formulation on the adiabatic temperature prediction	. The considered fresh gas composition corresponds to the inlet conditions of a premixed flame at equivalence ratio $\phi=0.8$, $T=500$ K, $P=1$ atm, with an increasing concentration of TiCl$_{4}$. When the TiCl$_{4}$ inlet concentration is increased, the effect of the model formulation becomes patent, as can be seen on Fig. \ref{fig:1dp_grionestep_phi08_temperature}, where the equilibrium temperatures corresponding to both model formulations are compared. When the inlet TiCl$_{4}$ mass fraction is 25\,\%, the equilibrium temperature is lowered down to 1681\,K, which represents a relative error of roughly 30\,\%. This demonstrates the importance of enthalpy conservation when high nanoparticles mass fractions are encountered, and the strong coupling between the particle and the gas phase under such conditions. Note that there is a break in the equilibrium temperature decrease around 25\,\% inlet TiCl$_{4}$ mass fraction. This is because increasing the TiCl$_{4}$ mass fraction artificially decreases the enthalpy. As the enthalpy is decreased, the equilibrium temperature decreases, but as the flame gets close to extinction then it cannot provide enough energy for the TiCl$_{4}$ conversion. This is why the non-conservative model underpredicts the TiO$_{2}$ mass fraction above 20\,\% inlet TiCl$_{4}$, as can be observed in Fig. \ref{fig:1dp_grionestep_phi08_ytitania}.

The effect on the species mass fractions is also significant, as it can be seen in Fig. \ref{fig:1dp_grionestep_phi08_yk}, where the equilibrium mass fractions -- which correspond to the mass fractions in the burnt gases in the 1D premixed flame -- computed using STANJAN \cite{reynolds_stanjan_1981} have been plotted. The mass fractions are ploteed here as they are the final quantity of interest for the experimenters. The titania species mass fractions remain unaffected as far as the inlet TiCl$_{4}$ mass fraction remains below 20\,\%. When further increasing the titania precursor mass fraction, the discrepancies between the conservative and non-conservative formulations become significant: the relative error can reach 20\,\% for TiO$_{2}$ and Cl$_{2}$, and 50\,\% for TiCl$_{4}$ mass fractions. Similar conclusions can be drawn from Fig. \ref{fig:1dp_grionestep_phi08_yflame}. The combustion products in the burnt gases are sensitive to the conservation of enthalpy as soon as $Y_{\text{TiCl}_{4}}$ reaches 10\,\%. The relative error in CO$_{2}$ mass fraction can then be about 45\,\%, the relative error in H$_{2}$O mass fraction can be of 10\,\%, and the relative error in CO mass fraction can reach 95\,\%.

In addition, neglecting the enthalpy of the particles can lead to severe numerical difficulties. The numerical method used for the present study is based on a full coupling of both phases. The set of discretized equations is solved by means of a modified Newton method. Therefore, the conservation of energy is critical. As already stated, when neglecting the particle enthalpy $\tilde{h}_{\text{p}}$, we could not obtain numerical convergence when TiCl$_{4}$ injected mass fraction was greater than $5 \%$, even though continuation techniques were employed. The system seems to become singular when the nanoparticle mass fraction becomes too large. This is due to the fact that with an implicit method when using the conservative model the condition $h(+\infty)=h(-\infty)$ is enforced. On the contrary, when neglecting $\tilde{h}_{\text{p}}$, the condition $\tilde{h}_{\text{g}}(+\infty) = \tilde{h}_{\text{g}}(-\infty)$ is imposed, which is not possible. If one uses an explicit or semi-implicit solver, such numerical difficulty is potentially circumvented, however the system of equations used is still non-conservative and may lead to converged but non-physical solutions.

As already stated, in premixed configurations the diffusion of nanoparticles is relatively negligible as the mixture rapidly reaches a thermodynamic equilibrium, especially here given the fast TiCl$_{4}$ conversion rates. For this reason, the computation of the diffusion velocity is not critical. However, in non-premixed counterflow configurations, relative diffusion due to concentration gradients and thermophoresis can play an important role in the flame and nanoparticle dynamics, as it will be discussed in the following section.

\section{Results in 1D non-premixed counterflow configuration}
\label{sec:1dcf}

In this section, we first study the importance of enthalpy conservation, then the importance of differential diffusion with respect to mass conservation. 1D non-premixed counterflow flames are calculated using the 1D counterflow model in the Regath code \cite{rodrigues_unsteady_2017, rodrigues_modelisation_2018}. In the configuration studied here, the low Mach number approximation applies.

\subsection{Enthalpy conservation}

\begin{figure}[H]
\begin{minipage}[t]{.5\textwidth}
 \begin{center}
\subfloat[Main species mass fractions and temperature.]{\includegraphics[height = 45mm, clip = true]{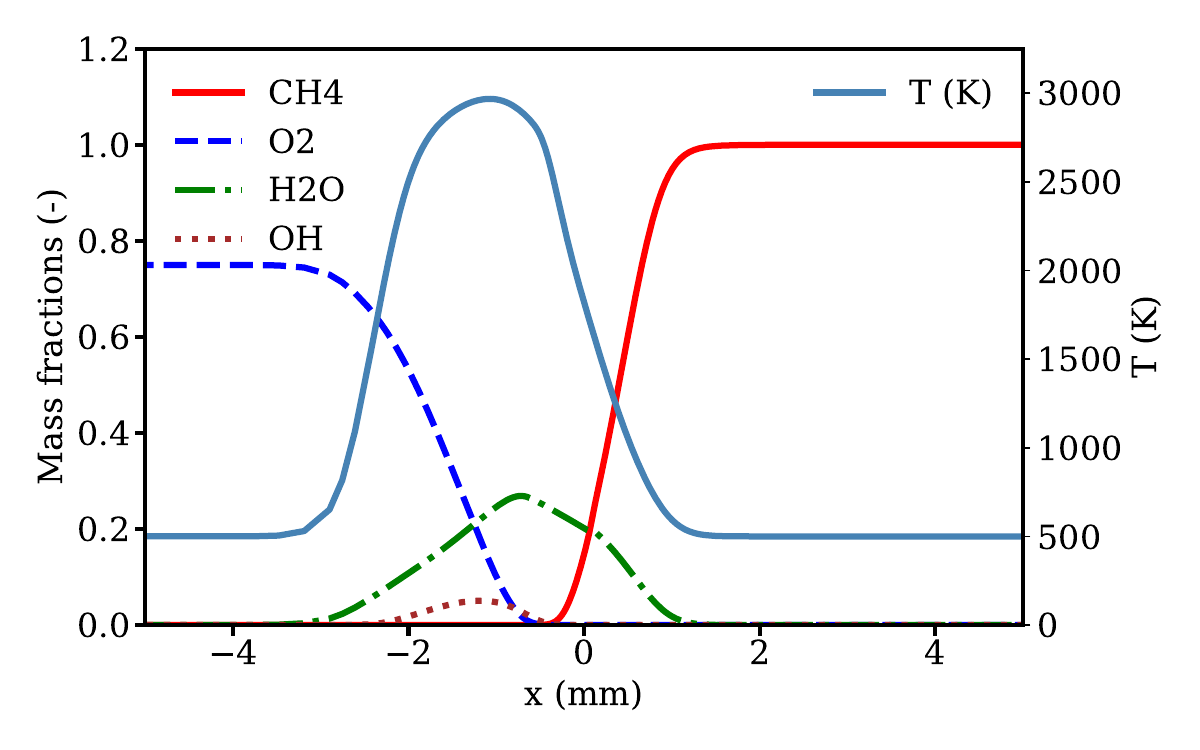}\label{fig:1dcf_grionestep_flame_ticl4_25e-2_alpha600}}
\end{center}
\end{minipage}%
\begin{minipage}[t]{.5\textwidth}
\begin{center}
\subfloat[Titania species mole fractions.]{\includegraphics[height = 45mm, clip = true]{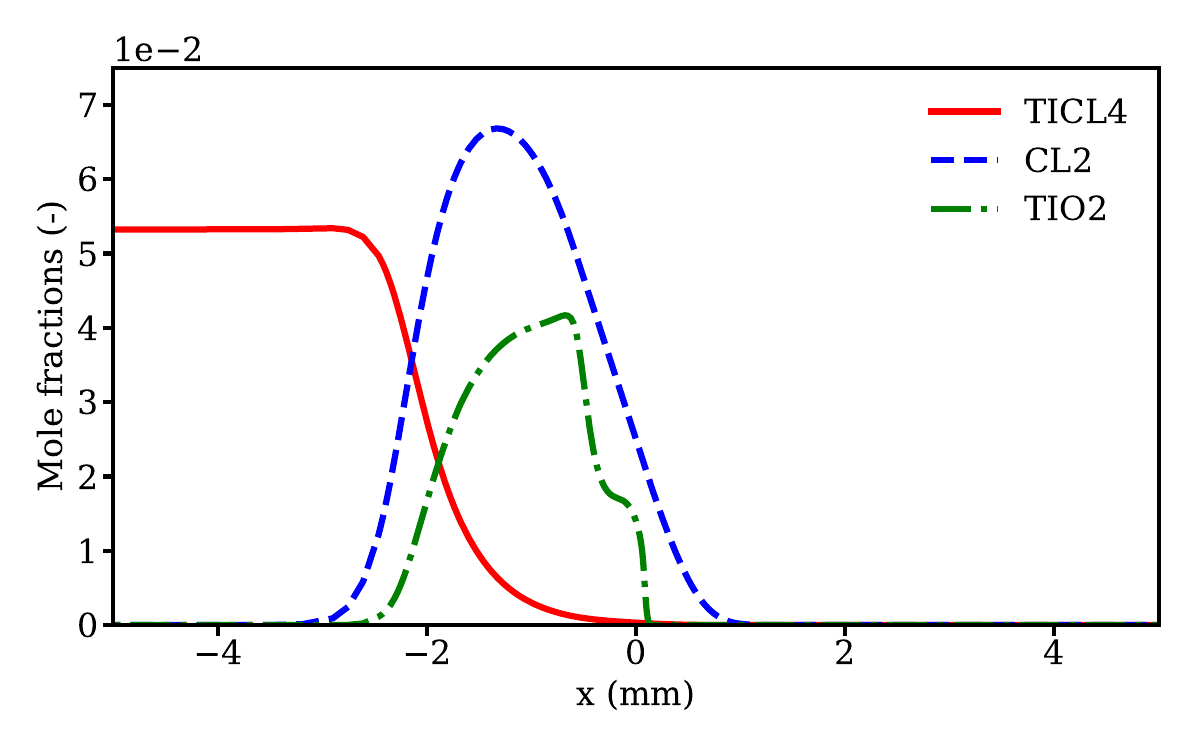}\label{fig:1dcf_grionestep_xtitania_ticl4_25e-2_alpha600}}
\end{center}
\end{minipage}
\caption{Main species profiles for the 1D counterflow CH$_{4}$/O$_{2}$ flame. The inlet temperature is 500~K. Injected TiCl$_{4}$ mass fraction $Y_{\text{TiCl}_{4}}^{\text{inj}}=0.25$ (oxidizer side: $x=-\infty$). Strain rate $\alpha=600 \,$s$^{-1}$. The stagnation plane is located at $x=0$\,mm. Results are obtained using the conservative model for mass and enthalpy.}
\label{fig:1dcf_grionestep_ticl4_25e-2_alpha600}
\end{figure}

We investigate here the importance of the global conservation of enthalpy in the mixture in a non-premixed CH$_{4}$/O$_{2}$ flame. Fig. \ref{fig:1dcf_grionestep_ticl4_25e-2_alpha600} shows results for a counterflow CH$_{4}$/O$_{2}$+TiCl$_{4}$ flame. The oxidizer mixture is injected from the left, and contains 25\,\% TiCl$_{4}$ and 75\,\% O$_{2}$ in mass, while the -- 100\,\% CH$_{4}$ -- fuel is injected from the right. The injection temperature is of 500\,K on both sides, and the strain rate is 600\,s$^{-1}$. The origin is set at the stagnation plane.

In Fig. \ref{fig:1dcf_grionestep_flame_ticl4_25e-2_alpha600}, respectively Fig. \ref{fig:1dcf_grionestep_xtitania_ticl4_25e-2_alpha600}, the main combustion species mass fraction and temperature profiles, respectively the titania species mole fraction profiles, are plotted. It can be seen that the maximum temperature and H$_{2}$O mass fraction -- Fig. \ref{fig:1dcf_grionestep_flame_ticl4_25e-2_alpha600} -- are located on the oxidizer side ($x<0$) due to high diffusion of CH$_{4}$. The TiCl$_{4}$ conversion into TiO$_{2}$ -- Fig. \ref{fig:1dcf_grionestep_xtitania_ticl4_25e-2_alpha600} -- is almost completed at maximum temperature while TiO$_{2}$ is formed as soon as H$_{2}$O mass fraction increases. The TiO$_{2}$ mole fraction has a non-linear behavior near the stagnation point because of the intricate effect of convection and thermophoresis. Indeed, the convection decreases as the nanoparticles get closer to the stagnation plane, while the thermophoretic force is first directed upstream of the flow, turns downstream as the nanoparticles cross the maximum temperature point, and then increases until the nanoparticles cross the zone of maximum temperature gradient.

\begin{figure}[H]
\begin{center}
\includegraphics[height = 60mm, clip = true]{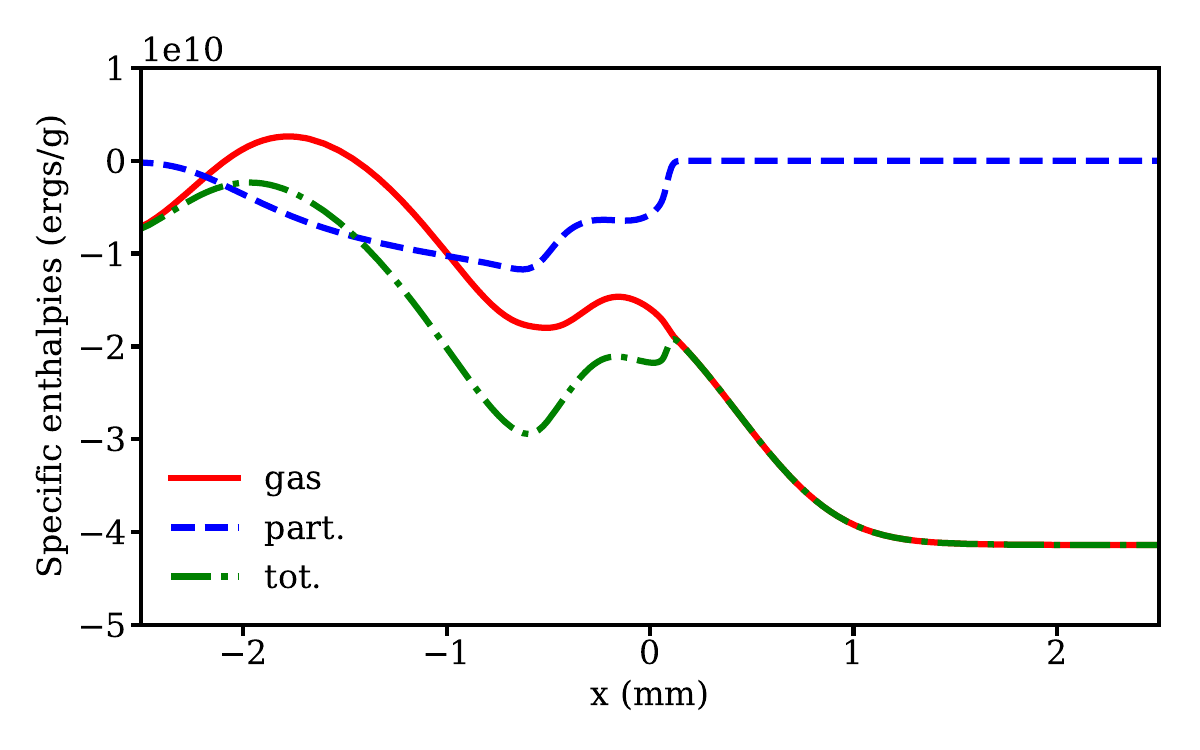}
\caption{Contributions $\tilde{h}_{\text{g}}$ and $\tilde{h}_{\text{p}}$ of the gas and solid particle phases to the mixture specific enthalpy $h$. Same conditions as in Fig. \ref{fig:1dcf_grionestep_ticl4_25e-2_alpha600}.}
\label{fig:1dcf_grionestep_hk_ticl4_25e-2_alpha600}
\end{center}
\end{figure}

In Fig. \ref{fig:1dcf_grionestep_hk_ticl4_25e-2_alpha600}, the enthalpies of the respective gas and particle phases are plotted for the 1D counterflow flame of Fig. \ref{fig:1dcf_grionestep_ticl4_25e-2_alpha600}. As in the premixed case, one can see that the enthalpy of the particle phase represents a non-negligible part of the enthalpy of the mixture in the counterflow flame. However, a much higher injected TiCl$_{4}$ mass fraction -- 25\,$\%$ -- is necessary compared to the premixed case -- 5\,$\%$ -- to observe a comparable relative contribution of TiO$_{2}$ enthalpy to the mixture enthalpy. This is because the absolute value of CH$_{4}$ specific enthalpy at $500$\,K is one order of magnitude larger than the specific enthalpies of O$_{2}$ and N$_{2}$, so that the relative contribution of TiO$_{2}$ appears lower. However, the absolute contribution in the counterflow flame -- approximately $-10^{10}$ ergs.g$^{-1}$ at $25\,\%$ TiCl$_{4}$ -- is consistent with the value observed in the premixed flame -- approximately $-2 \cdot 10^{9}$ ergs.g$^{-1}$ at $5\,\%$ TiCl$_{4}$.

It is worth noting that no convergence issues were encountered for this configuration when using the non-energy-conserving model. Indeed, while in the premixed flame the boundary conditions are constrained by the conservation of total enthalpy, in the counterflow flame no such constraint exists due to the lateral heat loss.

\begin{figure}[H]
\begin{center}
\includegraphics[height = 60mm, clip = true]{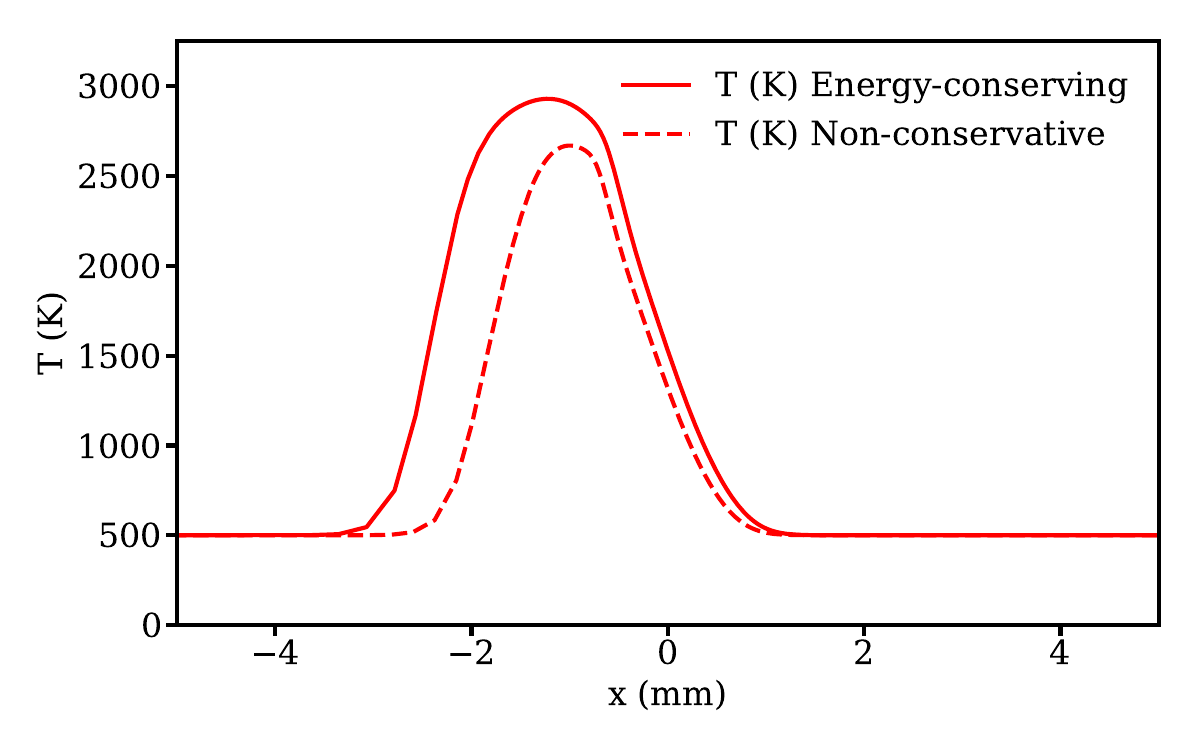}
\caption{Comparison of the temperature profiles obtained with the energy-conserving model (continuous line) and the non-energy-conserving one (dashed line). Same conditions as in Fig. \ref{fig:1dcf_grionestep_ticl4_25e-2_alpha600} with $0.5$ injected TiCl$_{4}$ mass fraction: $Y_{\text{TiCl}_{4}}^{\text{inj}}=0.5$.}
\label{fig:1dcf_grionestep_temperature_alpha600_ticl4_5e-1_Comp_Cons_nonCons}
\end{center}
\end{figure}

In Fig. \ref{fig:1dcf_grionestep_temperature_alpha600_ticl4_5e-1_Comp_Cons_nonCons} we compare the temperature profiles obtained with the energy-conserving model and the non-energy-conserving model -- see subsection \ref{subsec:enthalpy} -- while applying the mass-conserving model in both cases. As expected these profiles differ from each other in the region where nanoparticles are present. The differences in temperatures observed can locally reach 66\,\%. The width of the flame front is also not correctly predicted by the non-conservative model, which induces an artificial thinning of the flame.

\begin{figure}[H]
\begin{center}
\includegraphics[height = 60mm, clip = true]{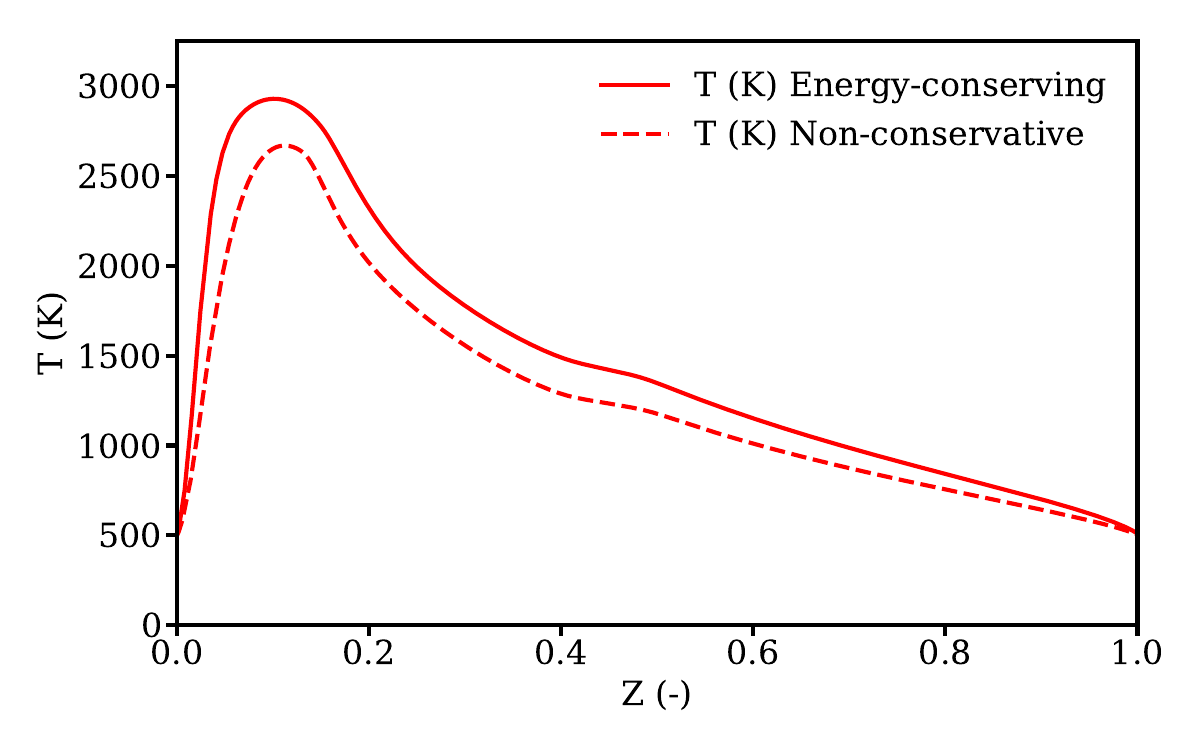}
\caption{Comparison of the temperature profiles obtained with the energy-conserving model (continuous line) and the non-energy-conserving one (dashed line). Same conditions as in Fig. \ref{fig:1dcf_grionestep_ticl4_25e-2_alpha600} with $0.5$ injected TiCl$_{4}$ mass fraction: $Y_{\text{TiCl}_{4}}^{\text{inj}}=0.5$.}
\label{fig:1dcf_grionestep_temperature_alpha600_ticl4_5e-1_Comp_Cons_nonCons_Z}
\end{center}
\end{figure}

In Fig. \ref{fig:1dcf_grionestep_temperature_alpha600_ticl4_5e-1_Comp_Cons_nonCons_Z} the same temperature profiles are plotted as a function of the mixture fraction $Z$. The flame structure is notably modified also in the mixture fraction space, as the maximum temperatures differ between the two formulations.

\begin{figure}[H]
\begin{minipage}[t]{.5\textwidth}
 \begin{center}
\subfloat[Ti-related species.]{\includegraphics[height = 45mm, clip = true]{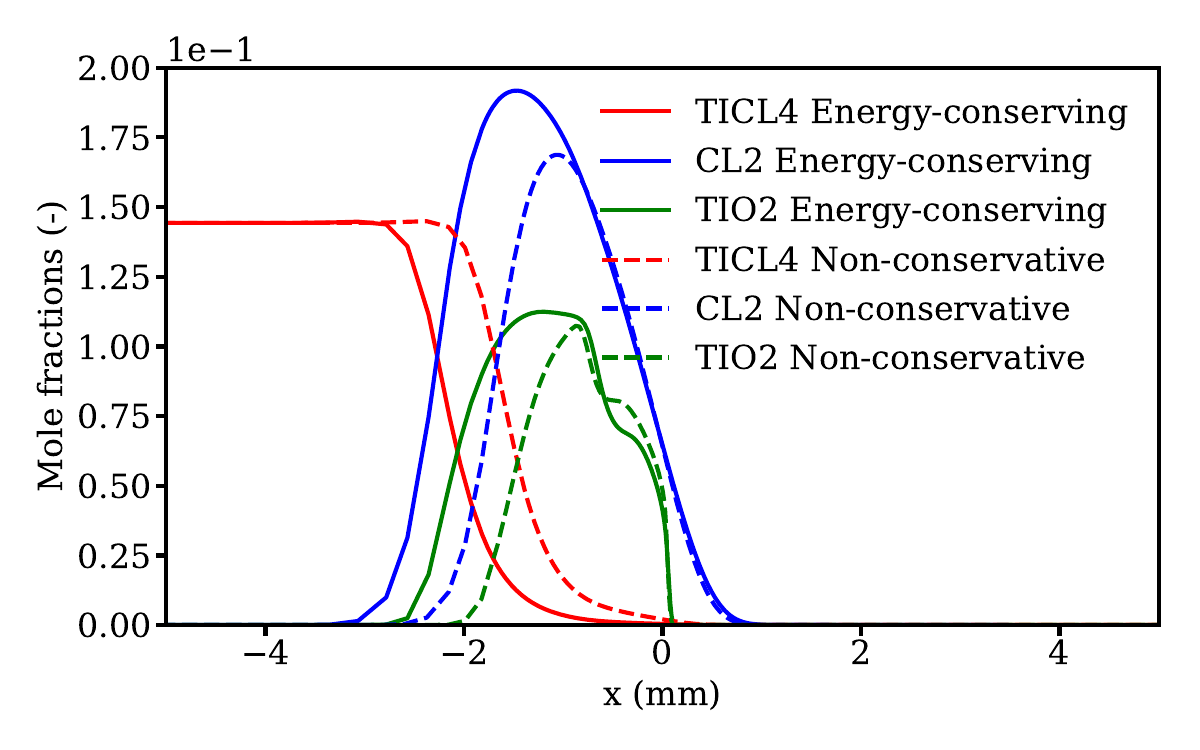}\label{fig:1dcf_grionestep_xtio2_alpha600_ticl4_5e-1_Comp_Cons_nonCons}}
\end{center}
\end{minipage}%
\begin{minipage}[t]{.5\textwidth}
\begin{center}
\subfloat[Combustion products.]{\includegraphics[height = 45mm, clip = true]{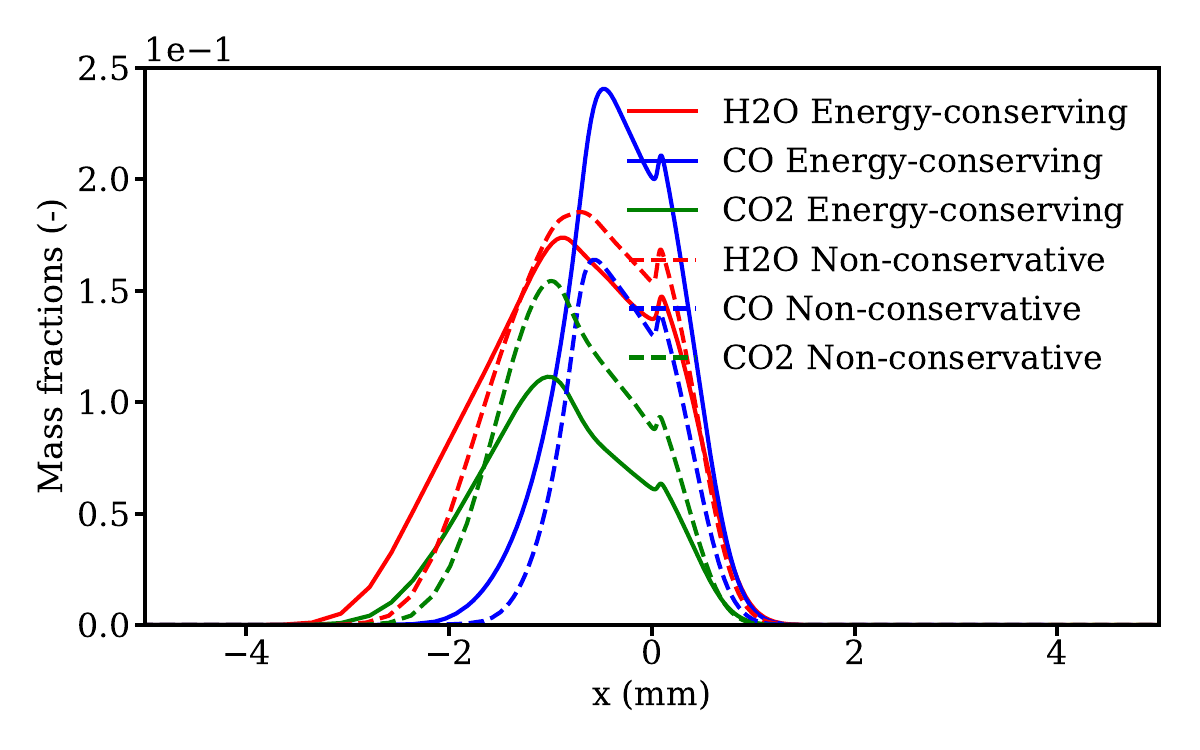}\label{fig:1dcf_grionestep_yh2o_alpha600_ticl4_5e-1_Comp_Cons_nonCons}}
\end{center}
\end{minipage}
\caption{Comparison of the species mass fraction profiles obtained with the energy-conserving (continuous lines) and the non-conservative (dashed lines) formulations in the counterflow flame. Same conditions as in Fig. \ref{fig:1dcf_grionestep_temperature_alpha600_ticl4_5e-1_Comp_Cons_nonCons}.}
\label{fig:1dcf_grionestep_ticl4_5e-1_alpha600_corgasseul}
\end{figure}

The TiCl$_{4}$, TiO$_{2}$ and Cl$_{2}$ mole fraction profiles, shown in Fig. \ref{fig:1dcf_grionestep_xtio2_alpha600_ticl4_5e-1_Comp_Cons_nonCons}, are also drastically affected by the conservation of enthalpy: the titania profile in particular is much narrower with the non-conservative model, which implies that the conversion yield is underpredicted. The combustion products, shown in Fig. \ref{fig:1dcf_grionestep_yh2o_alpha600_ticl4_5e-1_Comp_Cons_nonCons} are also affected, with about 13 \,\% relative error on H$_{2}$O mass fraction, and about 30\,\% relative error on CO$_{2}$ and CO mass fractions.

\subsection{Mass conservation}
\begin{figure}[H]
\begin{minipage}[t]{.5\textwidth}
 \begin{center}
\subfloat[TiCl$_{4}$ and TiO$_{2}$ mole fraction profiles.]{\includegraphics[height = 45mm, clip = true]{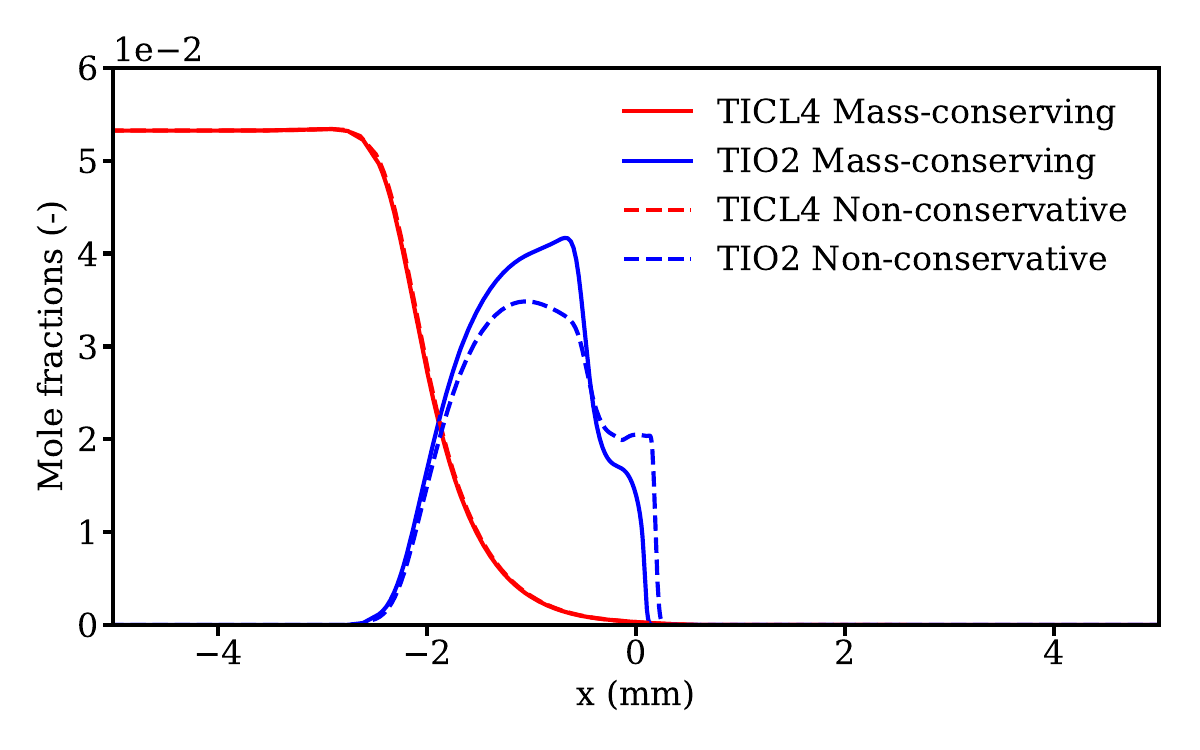}\label{fig:1dcf_grionestep_xtitania_ticl4_25e-2_alpha600_corgasseul}}
\end{center}
\end{minipage}%
\begin{minipage}[t]{.5\textwidth}
\begin{center}
\subfloat[Mass fraction profiles of H$_{2}$O, CO and CO$_{2}$.]{\includegraphics[height = 45mm, clip = true]{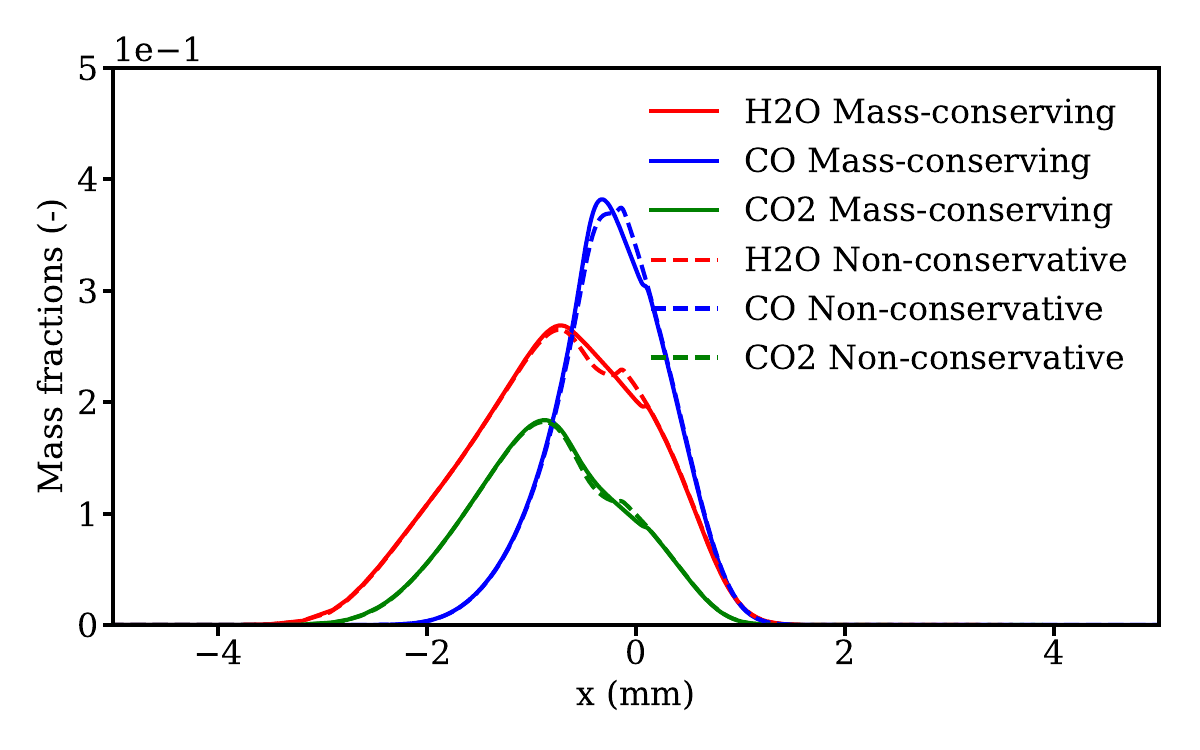}\label{fig:1dcf_grionestep_yh2o_ticl4_25e-2_alpha600_corgasseul}}
\end{center}
\end{minipage}
\caption{Comparison between the mass-conserving (continuous lines) and the non-conservative (dashed lines) formulations in the counterflow flame. Same conditions as in Fig. \ref{fig:1dcf_grionestep_ticl4_25e-2_alpha600}.}
\label{fig:1dcf_grionestep_ticl4_25e-2_alpha600_corgasseul}
\end{figure}

In 1D premixed configurations, the diffusion of particles plays a negligible role, while in counterflow flames one expects a more significant impact. To assess this impact, we run two calculations. In the first calculation, we use the conservative model, where the diffusion velocities are all adjusted so that the diffusive fluxes of both gaseous species and particles sum to zero, as in Eqs. \eqref{eq:OneMixtDiscDiffFluxGasCons} and \eqref{eq:OneMixtDiscDiffFluxPartCons}. This model is referred to as the {\og mass-conserving \fg} model. In a second calculation, the contribution of nanoparticles to the correction velocity is neglected, and only the gaseous species diffusion velocities are corrected, as in Eqs. \eqref{eq:TwoMixtDiscDiffFluxGas} and \eqref{eq:TwoMixtDiscDiffFluxPart}. The energy-conserving model is retained for both simulations. Fig. \ref{fig:1dcf_grionestep_ticl4_25e-2_alpha600_corgasseul} presents the comparison between the two calculations.

The differences in the titania mole fractions obtained with the two formulations are important. The non-conservative model induces a relative error on the TiO$_{2}$ mole fraction of up to 20\,\% compared to the conservative model. With the former, more nanoparticles cross the stagnation plane than with the latter, because thermophoresis is hindered by neutral drag. However, the effect on the flame structure appears negligible. Indeed, the temperature profiles -- not shown here -- are almost identical. The combustion products are slightly affected in the area where nanoparticles are present. The H$_{2}$O, CO and CO$_{2}$ mass fraction profiles, plotted in Fig. \ref{fig:1dcf_grionestep_yh2o_ticl4_25e-2_alpha600_corgasseul}, are very similar, excepted that the non-conservative model leads to the apparition of a local maximum in the H$_{2}$O and CO$_{2}$ profiles close to the stagnation plane, and a shift in the CO maximum mass fraction towards the stagnation plane.

In conclusion, the effect of using a non-conservative model can lead to large errors in both premixed and counterflow flames when a high concentration of nanoparticles is expected. When the particles enthalpy is neglected, the relative errors on the temperature profiles reaches 30\,\% in the considered premixed flame, and almost 70\,\% in the counterflow flame. The relative errors on the species mass fraction profiles can reach 95\,\%. Besides, the width of the counterflow flame is significantly affected and the global product yield is thus further underestimated. Also, the numerical stability may be compromised when using a non-conservative model. Using a conservative formulation is of first importance for accurate and detailed modeling of flame synthesis of nanoparticles for high particle mass fractions.

\section{Conclusion}
We have studied the modeling of titania nanoparticle production in flames with high product mass fraction. It has been shown that the conservation of enthalpy and mass are of great importance, both for numerical stability and physical consistency purposes.

The relative importance of the particle-phase enthalpy has been illustrated both in 1D premixed and in 1D counterflow calculations. It has been shown that the enthalpy of the particle phase can represent a significant part of the mixture enthalpy, which cannot be neglected, as often done in soot or fine particle models. It has further been shown that using a non-enthalpy-conserving scheme may yield significant errors in the temperature and mass fraction profiles for both premixed and counterflow flames. Additionally, using a non-energy-conserving formulation may lead to severe numerical difficulties.

Second, the importance of differential diffusion has been illustrated in 1D counterflow simulations. Even though the particles diffuse slower than the gaseous species, the conservation of mass requires to account for a correction velocity due to thermophoresis. Contrary to enthalpy conservation, no numerical instabilities have been observed due to non-conservation of mass. Yet, the formulation of the diffusion velocities has an important impact on the titania mass fraction profiles, though a marginal impact on the flame structure.

The importance of the formulation of the conservation equations has been therefore demonstrated on laminar flames, both for enthalpy and differential diffusion. For the enthalpy, the effect of the formulation on the flame structure is important, and can lead to large errors on the burnt gases flame temperature in premixed flames and on the flame width in counterflow flames. As a consequence, using a non-energy-conserving model can also have a strong impact on turbulent flames, as it can affect drastically the local flame structure. In a future study, the impact of using non-energy-conserving models in turbulent flames should be assessed.

As well, radiation has been neglected in this study, although radiation is known to influence deeply the flame and nanoparticles profiles, even at very low nanoparticles mass fraction. It is reasonable to expect that both phases remain in or close to thermal equilibrium between each other, and thus the effect of radiative heat loss would decrease the gas phase and the nanoparticle phase enthalpies concomitantly. Yet radiation would also have some significant effect on the chemical reaction rates, which still need to be evaluated. In a future work, the impact of radiation on the results presented here will be rigorously quantified.

Finally, the conservation of mass has been ensured here by means of a correction velocity, but it is well known that such a diffusion model is an approximation compared to full multicomponent diffusion models \cite{giovangigli_multicomponent_1999}. This justifies the need to derive a multicomponent diffusion model for reacting gases and nanoparticles mixtures. This can be done by extending existing methods from kinetic theory \cite{orlach_kinetic_2018}.

\section*{Funding}
The support of the European Research Council (ERC) under the European Union Horizon 2020 research and innovation programme (grant agreement No. 757912) is acknowledged.

\bibliographystyle{tfq}
\bibliography{sotuf}

\begin{thebibliography}{10}
\newcommand{\printfirst}[2]{#1}
\newcommand{\switchargs}[2]{#2#1}
\providecommand{\url}[1]{\normalfont{#1}}
\providecommand{\urlprefix}{Available at }

\bibitem{pratsinis_flame_1998}
E. Pratsinis, \emph{Flame aerosol synthesis of ceramic powders}, Prog. Energy
  Combust. Sci. 23 (1998), pp. 197--219.

\bibitem{xu_simultaneous_2015}
Z. Xu and H. Zhao, \emph{Simultaneous measurement of internal and external
  properties of nanoparticles in flame based on thermophoresis}, Combustion and
  Flame 162 (2015), pp. 2200--2213,
  \urlprefix\url{https://linkinghub.elsevier.com/retrieve/pii/S0010218015000309}.

\bibitem{kelesidis_flame_2017}
G.A. Kelesidis, E. Goudeli, and S.E. Pratsinis, \emph{Flame synthesis of
  functional nanostructured materials and devices: {Surface} growth and
  aggregation}, Proceedings of the Combustion Institute 36 (2017), pp. 29--50,
  \urlprefix\url{https://linkinghub.elsevier.com/retrieve/pii/S1540748916304679}.

\bibitem{schulz_gas-phase_2018}
C. Schulz, T. Dreier, M. Fikri, and H. Wiggers, \emph{Gas-phase synthesis of
  functional nanomaterials: {Challenges} to kinetics, diagnostics, and process
  development}, Proceedings of the Combustion Institute  (2018),
  \urlprefix\url{https://linkinghub.elsevier.com/retrieve/pii/S1540748918304176}.

\bibitem{weise_buoyancy_2013}
C. Weise, A. Faccinetto, S. Kluge, T. Kasper, H. Wiggers, C. Schulz, I. Wlokas,
  and A. Kempf, \emph{Buoyancy induced limits for nanoparticle synthesis
  experiments in horizontal premixed low-pressure flat-flame reactors},
  Combustion Theory and Modelling 17 (2013), pp. 504--521,
  \urlprefix\url{http://www.tandfonline.com/doi/abs/10.1080/13647830.2013.781224}.

\bibitem{raman_modeling_2016}
V. Raman and R.O. Fox, \emph{Modeling of {Fine}-{Particle} {Formation} in
  {Turbulent} {Flames}}, Annual Review of Fluid Mechanics 48 (2016), pp.
  159--190,
  \urlprefix\url{http://www.annualreviews.org/doi/10.1146/annurev-fluid-122414-034306}.

\bibitem{sellmann_detailed_2018}
J. Sellmann, I. Rahinov, S. Kluge, H. Jünger, A. Fomin, S. Cheskis, C. Schulz,
  H. Wiggers, A. Kempf, and I. Wlokas, \emph{Detailed simulation of iron oxide
  nanoparticle forming flames: {Buoyancy} and probe effects}, Proceedings of
  the Combustion Institute  (2018),
  \urlprefix\url{https://linkinghub.elsevier.com/retrieve/pii/S1540748918302244}.

\bibitem{george_formation_1973}
P. George, \emph{Formation of {Ti02} {Aerosol} from the {Combustion}
  {Supported} {Reaction} of {TiCl4} and {O2}}, Faraday Symposia of the Chemical
  Society 7 (1973), pp. 63--71.

\bibitem{pratsinis_role_1996}
S.E. Pratsinis, W. Zhu, and S. Vemury, \emph{The role of gas mixing in flame
  synthesis of titania powders}, Powder Technology 86 (1996), pp. 87--93,
  \urlprefix\url{http://linkinghub.elsevier.com/retrieve/pii/0032591095030417}.

\bibitem{chow_flame_1996}
W. Zhu and S.E. Pratsinis, \emph{Flame {Synthesis} of {Nanosize} {Powders}:
  {Effect} of {Flame} {Configuration} and {Oxidant} {Composition}}, in
  \emph{Nanotechnology}, G.M. Chow and K.E. Gonsalves, eds., {ACS} {Symposium}
  {Series} Vol. 622, American Chemical Society, Washington, DC,  1996, pp.
  64--78,
  \urlprefix\url{http://pubs.acs.org/doi/abs/10.1021/bk-1996-0622.ch004}.

\bibitem{pratsinis_kinetics_1990}
S.E. Pratsinis, H. Bai, P. Biswas, M. Frenklach, and S.V.R. Mastrangelo,
  \emph{Kinetics of {Titanium}({IV}) {Chloride} {Oxidation}}, Journal of the
  American Ceramic Society 73 (1990), pp. 2158--2162,
  \urlprefix\url{http://doi.wiley.com/10.1111/j.1151-2916.1990.tb05295.x}.

\bibitem{heine_agglomerate_2007}
M.C. Heine and S.E. Pratsinis, \emph{Agglomerate {TiO2} {Aerosol} {Dynamics} at
  {High} {Concentrations}}, Particle \& Particle Systems Characterization 24
  (2007), pp. 56--65,
  \urlprefix\url{http://doi.wiley.com/10.1002/ppsc.200601076}.

\bibitem{hwang_measurements_2001}
J.Y. Hwang, Y.S. Gil, J.I. Kim, M. Choi, and S.H. Chung, \emph{Measurements of
  temperature and {OH} radical distributions in a silica generating \#ame using
  {CARS} and {PLIF}}, Aerosol Science  (2001), p.~13.

\bibitem{kim_modeling_2003}
H.J. Kim, J.I. Jeong, Y. Park, Y. Yoon, and M. Choi, \emph{Modeling of
  {Generation} and {Growth} of {Non}-{Spherical} {Nanoparticles} in a
  {Co}-{Flow} {Flame}}, Journal of Nanoparticle Research 5 (2003), pp.
  237--246.

\bibitem{lee_simulation_2001}
B. Lee, S. Oh, and M. Choi, \emph{Simulation of {Growth} of {Nonspherical}
  {Silica} {Nanoparticles} in a {Premixed} {Flat} {Flame}}, Aerosol Science and
  Technology 35 (2001), pp. 978--989,
  \urlprefix\url{http://www.tandfonline.com/doi/abs/10.1080/027868201753306741}.

\bibitem{morgan_modelling_2005}
N. Morgan, C. Wells, M. Kraft, and W. Wagner, \emph{Modelling nanoparticle
  dynamics: coagulation, sintering, particle inception and surface growth},
  Combustion Theory and Modelling 9 (2005), pp. 449--461,
  \urlprefix\url{http://www.tandfonline.com/doi/abs/10.1080/13647830500277183}.

\bibitem{grohn_design_2011}
A.J. Gröhn, B. Buesser, J.K. Jokiniemi, and S.E. Pratsinis, \emph{Design of
  {Turbulent} {Flame} {Aerosol} {Reactors} by {Mixing}-{Limited} {Fluid}
  {Dynamics}}, Industrial \& Engineering Chemistry Research 50 (2011), pp.
  3159--3168, \urlprefix\url{http://pubs.acs.org/doi/abs/10.1021/ie1017817}.

\bibitem{chrystie_comparative_2018}
R.S. Chrystie, H. Janbazi, T. Dreier, H. Wiggers, I. Wlokas, and C. Schulz,
  \emph{Comparative study of flame-based {SiO2} nanoparticle synthesis from
  {TMS} and {HMDSO}: {SiO}-{LIF} concentration measurement and detailed
  simulation}, Proceedings of the Combustion Institute  (2018),
  \urlprefix\url{https://linkinghub.elsevier.com/retrieve/pii/S1540748918304425}.

\bibitem{pratsinis_competition_1998}
S.E. Pratsinis and P.T. Spicer, \emph{Competition between gas phase and surface
  oxidation of {TiCl4} during synthesis of {TiO2} particles}, Chemical
  Engineering Science 53 (1998), pp. 1861--1868,
  \urlprefix\url{http://linkinghub.elsevier.com/retrieve/pii/S0009250998000268}.

\bibitem{spicer_titania_2002}
P.T. Spicer, O. Chaoul, S. Tsantilis, and S.E. Pratsinis, \emph{Titania
  formation by {TiCl4} gas phase oxidation, surface growth and coagulation},
  Journal of Aerosol Science 33 (2002), pp. 17--34,
  \urlprefix\url{http://linkinghub.elsevier.com/retrieve/pii/S0021850201000696}.

\bibitem{tsantilis_narrowing_2004}
S. Tsantilis and S.E. Pratsinis, \emph{Narrowing the size distribution of
  aerosol-made titania by surface growth and coagulation}, Journal of Aerosol
  Science 35 (2004), pp. 405--420,
  \urlprefix\url{http://linkinghub.elsevier.com/retrieve/pii/S0021850203004415}.

\bibitem{morgan_new_2006}
N. Morgan, C. Wells, M. Goodson, M. Kraft, and W. Wagner, \emph{A new numerical
  approach for the simulation of the growth of inorganic nanoparticles},
  Journal of Computational Physics 211 (2006), pp. 638--658,
  \urlprefix\url{http://linkinghub.elsevier.com/retrieve/pii/S0021999105002913}.

\bibitem{boje_detailed_2017}
A. Boje, J. Akroyd, S. Sutcliffe, J. Edwards, and M. Kraft, \emph{Detailed
  population balance modelling of {TiO2} synthesis in an industrial reactor},
  Chemical Engineering Science 164 (2017), pp. 219--231,
  \urlprefix\url{https://linkinghub.elsevier.com/retrieve/pii/S0009250917301276}.

\bibitem{west_detailed_2009}
R.H. West, R.A. Shirley, M. Kraft, C.F. Goldsmith, and W.H. Green, \emph{A
  detailed kinetic model for combustion synthesis of titania from {TiCl4}},
  Combustion and Flame 156 (2009), pp. 1764--1770,
  \urlprefix\url{http://linkinghub.elsevier.com/retrieve/pii/S0010218009001163}.

\bibitem{mehta_multiscale_2010}
M. Mehta, Y. Sung, V. Raman, and R.O. Fox, \emph{Multiscale {Modeling} of
  {TiO2} {Nanoparticle} {Production} in {Flame} {Reactors}: {Effect} of
  {Chemical} {Mechanism}}, Industrial \& Engineering Chemistry Research 49
  (2010), pp. 10663--10673,
  \urlprefix\url{http://pubs.acs.org/doi/abs/10.1021/ie100560h}.

\bibitem{mehta_role_2013}
M. Mehta, V. Raman, and R.O. Fox, \emph{On the role of gas-phase and surface
  chemistry in the production of titania nanoparticles in turbulent flames},
  Chemical Engineering Science 104 (2013), pp. 1003--1018,
  \urlprefix\url{http://linkinghub.elsevier.com/retrieve/pii/S0009250913007264}.

\bibitem{johannessen_computational_2001}
T. Johannessen, S.E. Pratsinis, and H. Livbjerg, \emph{Computational analysis
  of coagulation and coalescence in the flame synthesis of titania particles},
  Powder Technology 118 (2001), pp. 242--250,
  \urlprefix\url{http://linkinghub.elsevier.com/retrieve/pii/S0032591000004010}.

\bibitem{wang_modeling_2005}
G. Wang and S.C. Garrick, \emph{Modeling and {Simulation} of {Titania}
  {Synthesis} in {Two}-dimensional {Methane}–air {Flames}}, Journal of
  Nanoparticle Research 7 (2005), pp. 621--632,
  \urlprefix\url{http://link.springer.com/10.1007/s11051-005-4966-7}.

\bibitem{garrick_modeling_2011}
S.C. Garrick and G. Wang, \emph{Modeling and simulation of titanium dioxide
  nanoparticle synthesis with finite-rate sintering in planar jets}, Journal of
  Nanoparticle Research 13 (2011), pp. 973--984,
  \urlprefix\url{http://link.springer.com/10.1007/s11051-010-0097-x}.

\bibitem{akroyd_coupled_2011}
J. Akroyd, A.J. Smith, R. Shirley, L.R. McGlashan, and M. Kraft, \emph{A
  coupled {CFD}-population balance approach for nanoparticle synthesis in
  turbulent reacting flows}, Chemical Engineering Science 66 (2011), pp.
  3792--3805,
  \urlprefix\url{http://linkinghub.elsevier.com/retrieve/pii/S0009250911003009}.

\bibitem{xu_cfd-population_2017}
Z. Xu, H. Zhao, and H. Zhao, \emph{{CFD}-population balance {Monte} {Carlo}
  simulation and numerical optimization for flame synthesis of {TiO2}
  nanoparticles}, Proceedings of the Combustion Institute 36 (2017), pp.
  1099--1108,
  \urlprefix\url{https://linkinghub.elsevier.com/retrieve/pii/S1540748916302644}.

\bibitem{zimmer_investigation_2017}
L. Zimmer, F.M. Pereira, J.A. van  Oijen, and L.P.H. de  Goey,
  \emph{Investigation of mass and energy coupling between soot particles and
  gas species in modelling ethylene counterflow diffusion flames}, Combustion
  Theory and Modelling 21 (2017), pp. 358--379,
  \urlprefix\url{https://www.tandfonline.com/doi/full/10.1080/13647830.2016.1238512}.

\bibitem{smooke_investigation_2004}
M.D. Smooke, R.J. Hall, M.B. Colket~4, J. Fielding, M.B. Long, C.S. McEnally,
  and L.D. Pfefferle, \emph{Investigation of the transition from lightly
  sooting towards heavily sooting co-flow ethylene diffusion flames},
  Combustion Theory and Modelling 8 (2004), pp. 593--606,
  \urlprefix\url{https://www.tandfonline.com/doi/full/10.1088/1364-7830/8/3/009}.

\bibitem{wang_formation_2011}
H. Wang, \emph{Formation of nascent soot and other condensed-phase materials in
  flames}, Proceedings of the Combustion Institute 33 (2011), pp. 41--67,
  \urlprefix\url{http://linkinghub.elsevier.com/retrieve/pii/S1540748910003937}.

\bibitem{franzelli_time-resolved_2015}
B. Franzelli, P. Scouflaire, and S. Candel, \emph{Time-resolved spatial
  patterns and interactions of soot, {PAH} and {OH} in a turbulent diffusion
  flame}, Proceedings of the Combustion Institute 35 (2015), pp. 1921--1929,
  \urlprefix\url{https://linkinghub.elsevier.com/retrieve/pii/S1540748914002818}.

\bibitem{eaves_coflame:_2016}
N.A. Eaves, Q. Zhang, F. Liu, H. Guo, S.B. Dworkin, and M.J. Thomson,
  \emph{{CoFlame}: {A} refined and validated numerical algorithm for modeling
  sooting laminar coflow diffusion flames}, Computer Physics Communications 207
  (2016), pp. 464--477,
  \urlprefix\url{https://linkinghub.elsevier.com/retrieve/pii/S0010465516301813}.

\bibitem{mueller_joint_2009}
M.E. Mueller, G. Blanquart, and H. Pitsch, \emph{A joint volume-surface model
  of soot aggregation with the method of moments}, Proceedings of the
  Combustion Institute 32 (2009), pp. 785--792,
  \urlprefix\url{http://linkinghub.elsevier.com/retrieve/pii/S1540748908003313}.

\bibitem{bisetti_formation_2012}
F. Bisetti, G. Blanquart, M.E. Mueller, and H. Pitsch, \emph{On the formation
  and early evolution of soot in turbulent nonpremixed flames}, Combustion and
  Flame 159 (2012), pp. 317--335,
  \urlprefix\url{https://linkinghub.elsevier.com/retrieve/pii/S0010218011001672}.

\bibitem{xiong_formation_1993}
Y. Xiong, M. Kamal~Akhtar, and S.E. Pratsinis, \emph{Formation of agglomerate
  particles by coagulation and sintering—{Part} {II}. {The} evolution of the
  morphology of aerosol-made titania, silica and silica-doped titania powders},
  Journal of Aerosol Science 24 (1993), pp. 301--313,
  \urlprefix\url{http://linkinghub.elsevier.com/retrieve/pii/002185029390004S}.

\bibitem{buesser_coagulation_2009}
B. Buesser, M. Heine, and S. Pratsinis, \emph{Coagulation of highly
  concentrated aerosols}, Journal of Aerosol Science 40 (2009), pp. 89--100,
  \urlprefix\url{http://linkinghub.elsevier.com/retrieve/pii/S0021850208001730}.

\bibitem{friedlander_smoke_2000}
S. Friedlander and P. Friedlander, \emph{Smoke, {Dust}, and {Haze}:
  {Fundamentals} of {Aerosol} {Dynamics}}, Topics in chemical engineering,
  Oxford University Press, 2000,
  \urlprefix\url{https://books.google.fr/books?id=fNIeNvd3Ch0C}.

\bibitem{giovangigli_mass_1990}
V. Giovangigli, \emph{Mass {Conservation} and {Singular} {Multicomponent}
  {Diffusion} {Algorithms}}, Impact of Computing in Science and Engineering 2
  (1990), pp. 73--97.

\bibitem{gelbard_sectional_1980}
F. Gelbard, Y. Tambour, and J.H. Seinfeld, \emph{Sectional representations for
  simulating aerosol dynamics}, Journal of Colloid and Interface Science 76
  (1980), pp. 541--556,
  \urlprefix\url{https://linkinghub.elsevier.com/retrieve/pii/002197978090394X}.

\bibitem{derjaguin_experimental_1966}
B.V. Derjaguin, A.I. Storozhilova, and Y.I. Rabinovich, \emph{Experimental
  verification of the theory of thermophoresis of aerosol particles}, Journal
  of Colloid and Interface Science 21 (1966), pp. 35--58.

\bibitem{epstein_resistance_1924}
P.S. Epstein, \emph{On the {Resistance} {Experienced} by {Spheres} in their
  {Motion} through {Gases}}, Physical Review 23 (1924), pp. 710--733,
  \urlprefix\url{https://link.aps.org/doi/10.1103/PhysRev.23.710}.

\bibitem{waldmann_thermophoresis_1966}
L. Waldmann and K. Schmitt, \emph{Thermophoresis and diffusiophoresis of
  aerosols}, in \emph{Aerosol {Science}}, davies, c. n. ed., Academic Press,
  New York,  1966, pp. 137--162.

\bibitem{zhao_measurement_2003}
B. Zhao, Z. Yang, M.V. Johnston, H. Wang, A.S. Wexler, M. Balthasar, and M.
  Kraft, \emph{Measurement and numerical simulation of soot particle size
  distribution functions in a laminar premixed ethylene-oxygen-argon flame},
  Combustion and Flame 133 (2003), pp. 173--188,
  \urlprefix\url{https://linkinghub.elsevier.com/retrieve/pii/S0010218002005746}.

\bibitem{wang_experiments_1996}
H. Wang, D. Du, C. Sung, and C. Law, \emph{Experiments and numerical simulation
  on soot formation in opposed-jet ethylene diffusion flames}, Symposium
  (International) on Combustion 26 (1996), pp. 2359--2368,
  \urlprefix\url{https://linkinghub.elsevier.com/retrieve/pii/S0082078496800658}.

\bibitem{attili_formation_2014}
A. Attili, F. Bisetti, M.E. Mueller, and H. Pitsch, \emph{Formation, growth,
  and transport of soot in a three-dimensional turbulent non-premixed jet
  flame}, Combustion and Flame 161 (2014), pp. 1849--1865,
  \urlprefix\url{https://linkinghub.elsevier.com/retrieve/pii/S0010218014000133}.

\bibitem{jocher_impact_2017}
A. Jocher, K.K. Foo, Z. Sun, B. Dally, H. Pitsch, Z. Alwahabi, and G. Nathan,
  \emph{Impact of acoustic forcing on soot evolution and temperature in
  ethylene-air flames}, Proceedings of the Combustion Institute 36 (2017), pp.
  781--788,
  \urlprefix\url{https://linkinghub.elsevier.com/retrieve/pii/S154074891630414X}.

\bibitem{rodrigues_unsteady_2017}
P. Rodrigues, B. Franzelli, R. Vicquelin, O. Gicquel, and N. Darabiha,
  \emph{Unsteady dynamics of {PAH} and soot particles in laminar counterflow
  diffusion flames}, Proceedings of the Combustion Institute 36 (2017), pp.
  927--934,
  \urlprefix\url{https://linkinghub.elsevier.com/retrieve/pii/S1540748916303054}.

\bibitem{smith_gri-mech_nodate}
G.P. Smith, D.M. Golden, M. Frenklach, N.W. Moriarty, B. Eiteneer, M.
  Goldenberg, C.T. Bowman, R.K. Hanson, S. Song, W.C. Gardiner Jr., V.V.
  Lissianski, and Z. Qin, \emph{{GRI}-{Mech} 3.0}.
  \urlprefix\url{http://www.me.berkeley.edu/gri_mech/}.

\bibitem{mehta_reduced_2015}
M. Mehta, R.O. Fox, and P. Pepiot, \emph{Reduced {Chemical} {Kinetics} for the
  {Modeling} of {TiO2} {Nanoparticle} {Synthesis} in {Flame} {Reactors}},
  Industrial \& Engineering Chemistry Research 54 (2015), pp. 5407--5415,
  \urlprefix\url{http://pubs.acs.org/doi/10.1021/acs.iecr.5b00130}.

\bibitem{marchal_modelisation_2008}
C. Marchal, \emph{Modélisation de la formation et de l'oxydation des suies
  dans un moteur automobile}, Ph.D. diss., Université d'Orléans,  2008.

\bibitem{rodrigues_modelisation_2018}
P. Rodrigues, \emph{Modélisation multiphysique de flammes turbulentes suitées
  avec la prise en compte des transferts radiatifs et des transferts de chaleur
  pariétaux.}, Ph.D. diss., Université Paris-Saclay,  2018.

\bibitem{sutherland_viscosity_1893}
W. Sutherland, \emph{The viscosity of gases and molecular force}, The London,
  Edinburgh, and Dublin Philosophical Magazine and Journal of Science 36
  (1893), pp. 507--531,
  \urlprefix\url{https://www.tandfonline.com/doi/full/10.1080/14786449308620508}.

\bibitem{cunningham_velocity_1910}
E. Cunningham, \emph{On the {Velocity} of {Steady} {Fall} of {Spherical}
  {Particles} through {Fluid} {Medium}}, Proceedings of the Royal Society A:
  Mathematical, Physical and Engineering Sciences 83 (1910), pp. 357--365,
  \urlprefix\url{http://rspa.royalsocietypublishing.org/cgi/doi/10.1098/rspa.1910.0024}.

\bibitem{pratsinis_simultaneous_1988}
S.E. Pratsinis, \emph{Simultaneous nucleation, condensation, and coagulation in
  aerosol reactors}, Journal of Colloid and Interface Science 124 (1988), pp.
  416--427,
  \urlprefix\url{http://linkinghub.elsevier.com/retrieve/pii/0021979788901804}.

\bibitem{ghoshtagore_mechanism_1970}
R.N. Ghoshtagore, \emph{Mechanism of {Heterogeneous} {Deposition} of {Thin}
  {Film} {Rutile}}, Journal of The Electrochemical Society 117 (1970), p. 529,
  \urlprefix\url{http://jes.ecsdl.org/cgi/doi/10.1149/1.2407561}.

\bibitem{shirley_theoretical_2011}
R. Shirley, J. Akroyd, L.A. Miller, O.R. Inderwildi, U. Riedel, and M. Kraft,
  \emph{Theoretical insights into the surface growth of rutile {TiO2}},
  Combustion and Flame 158 (2011), pp. 1868--1876,
  \urlprefix\url{http://linkinghub.elsevier.com/retrieve/pii/S0010218011001878}.

\bibitem{nakaso_size_2001}
K. Nakaso, T. Fujimoto, T. Seto, M. Shimada, K. Okuyama, and M.M. Lunden,
  \emph{Size {Distribution} {Change} of {Titania} {Nano}-{Particle}
  {Agglomerates} {Generated} by {Gas} {Phase} {Reaction}, {Agglomeration}, and
  {Sintering}}, Aerosol Science and Technology 35 (2001), pp. 929--947,
  \urlprefix\url{http://www.tandfonline.com/doi/abs/10.1080/02786820126857}.

\bibitem{west_first-principles_2007}
R.H. West, G.J.O. Beran, W.H. Green, and M. Kraft, \emph{First-{Principles}
  {Thermochemistry} for the {Production} of {TiO} $_{\textrm{2}}$ from {TiCl}
  $_{\textrm{4}}$}, The Journal of Physical Chemistry A 111 (2007), pp.
  3560--3565, \urlprefix\url{http://pubs.acs.org/doi/abs/10.1021/jp0661950}.

\bibitem{reynolds_stanjan_1981}
W.C. Reynolds, \emph{{STANJAN}: {Interactive} {Computer} {Programs} for
  {Chemical} {Equilibrium} {Analysis}}, Tech. {R}ep., Stanford University Dept
  of Mechanical Engineering Thermosciences,  1981.

\bibitem{giovangigli_multicomponent_1999}
V. Giovangigli, \emph{Multicomponent {Flow} {Modeling}}, Birkhäuser, 1999.

\bibitem{orlach_kinetic_2018}
J.M. Orlac’h, V. Giovangigli, T. Novikova, and P. Roca i Cabarrocas,
  \emph{Kinetic theory of two-temperature polyatomic plasmas}, Physica A:
  Statistical Mechanics and its Applications 494 (2018), pp. 503--546,
  \urlprefix\url{https://linkinghub.elsevier.com/retrieve/pii/S0378437117312323}.

\end{thebibliography}

\end{document}